\documentclass[10pt]{article}

\usepackage{natbib}
\usepackage{times}
\usepackage{fancyhdr}
\usepackage{latexsym,amsmath}
\usepackage{amsmath}
\usepackage[pdftex]{graphicx}
\usepackage{multirow}
\usepackage{epstopdf}
\usepackage{amssymb}
\usepackage{epsfig}
\usepackage{xcolor}
\usepackage[normalem]{ulem}
\usepackage{url}
\usepackage{subfigure}

\definecolor{g}{rgb}{0.8, 0., 0.}
\definecolor{mas}{rgb}{0.54, 0.1, 0.95}

\newenvironment{sciabstract}{%
\begin{quote} \bf}
{\end{quote}}

\title{An anomalous topological phase transition in spatial random graphs}

\author
{Jasper van der Kolk$^{1,2}$, M. \'Angeles Serrano$^{1,2,3}$, Mari\'an Bogu\~n\'a$^{1,2\ast}$ \\
\\
\normalsize{$^{1}$Departament de F{\'\i}sica de la Mat\`eria Condensada, Universitat de Barcelona,}\\
\normalsize{Mart\'{\i} i Franqu\`es 1, 08028 Barcelona, Spain}\\
\normalsize{$^{2}$Universitat de Barcelona Institute of Complex Systems (UBICS),}\\
\normalsize{Universitat de Barcelona, Barcelona, Spain}\\
\normalsize{$^{3}$ICREA, Pg. Llu\'is Companys 23, E-08010 Barcelona, Spain}\\
\\
\normalsize{$^\ast$Corresponding author E-mail: marian.boguna@ub.edu (M.B.)}
}

\date{}

\begin{document} 


\baselineskip24pt


\maketitle 


\begin{sciabstract}
Clustering--the tendency for neighbors of nodes to be connected--quantifies the coupling of a complex network to its latent metric space. In random geometric graphs, clustering undergoes a continuous phase transition, separating a phase with finite clustering from a regime where clustering vanishes in the thermodynamic limit. We prove this geometric-to-nongeometric phase transition to be topological in nature, with anomalous features such as diverging entropy as well as atypical finite size scaling behavior of clustering. Moreover, a slow decay of clustering in the nongeometric phase implies that some real networks with relatively high levels of clustering may be better described in this regime.
\end{sciabstract}

\maketitle

\section*{Introduction}

For many years, Landau's theory of symmetry breaking was believed to be the ultimate explanation of continuous phase transitions~\cite{HOHENBERG20151}. In the liquid-crystal transition, for instance, the continuous translational and rotational symmetry at high temperatures break into a set of discrete symmetries in the low temperature phase. This paradigm was challenged for the first time by Berezinskii, Kosterlitz, and Thouless (BKT) in the two dimensional XY model~\cite{berezinskii1971destruction,berezinskii1972destruction,Kosterlitz_1973}. For this model, the Mermin-Wanger theorem~\cite{Mermin:1966fk} states that there is no ordered phase even at zero temperature, so that a phase transition in Landau's sense cannot exist. Yet, BKT showed that, in fact, there is a finite temperature phase transition driven by topological defects: vortices and antivortices. At low temperature, vortex-antivortex pairs are bound together. Above the critical temperature, vortex-antivortex pairs unbind, moving freely on the surface. No symmetry is broken in the transition since both phases are rotationally invariant and so magnetization is zero in both phases. Topological order and topological phase transitions are nowadays fundamental to understand the properties of quantum matter~\cite{Chiu:2016uq}.

In this paper, we show that a transition taking place in a very general class of sparse spatial random networks models is, in fact, topological in nature with no broken symmetry. In this transition, chordless cycles in the network play the role of topological defects with respect to a tree. A critical temperature separates a low temperature phase, where the underlying metric space forces chordless cycles to be short range --mostly triangles-- and a high temperature phase, where chordless cycles decouple from the metric space and become of the order of the network diameter. This is similar to the unbinding of vortex-antivortex pairs in the BKT transition. However, the thermodynamics of the transition is very different. As opposed to the BKT transition, the entropy density diverges at the critical temperature. This is also at odds with the continuous entropy density (with discontinuous first--or higher order--derivative) usually observed in continuous phase transitions. We thus describe a topological phase transition with novel thermodynamic properties. The two distinct topological orders of the transition can be quantified by means of the average local clustering coefficient, a measure of the fraction of triangles attached to nodes. Clustering is finite in the ``geometric'' phase with short range cycles --as a result of the triangle inequality of the underlying metric space-- and vanishes in the thermodynamic limit of the ``non-geometric'' phase with long range chordless cycles. This geometric-to-nongeometric phase transition shows interesting atypical scaling behavior as compared with standard continuous phase transitions, where one observes a power law decay at the critical point and a faster decay in the disordered phase. Instead, at the critical point, the average local clustering coefficient decays logarithmically to zero for very large systems and, in the nongeometric phase, where the coefficient decays as a power law, we discover a {\it quasi-geometric} region where the exponent that characterizes this decay depends on the temperature. 

\section*{Results and discussion}

We use a geometric description of networks~\cite{boguna2020network}, which provides a simple and comprehensive approach to complex networks. The existence of latent metric spaces underlying complex networks offers a deft explanation for their intricate topologies, giving at the same time important clues on their functionality. The small-world property, high levels of clustering, heterogeneity in the degree distribution, and hierarchical organization are all topological properties observed in real networks that find a simple explanation within the network geometry paradigm~\cite{boguna2020network}. Within this paradigm, the results found in this work hold in a very general class of spatial networks defined in compact homogeneous and isotropic Riemannian manifolds of arbitrary dimensionality~\cite{Boguna:2020fj,serrano2008similarity,bringmann2019geometric,kosmidis2008structural,Biskup:2004fk,Millan:2021uq}. Yet, in this paper, we focus on the $\mathbb{S}^1$ model~\cite{serrano2008similarity} and its isomorphically equivalent formulation in the hyperbolic plane, the $\mathbb{H}^2$ model~\cite{krioukov2010hyperbolic}. Interestingly, many analytic results have been derived for the 
$\mathbb{S}^1/\mathbb{H}^2$ model, e.~g. degree distribution~\cite{serrano2008similarity,krioukov2010hyperbolic,gugelmann2012random}, clustering~\cite{krioukov2010hyperbolic,gugelmann2012random,candellero2016clustering,Fountoulakis2021}, diameter~\cite{abdullah2017typical,friedrich2018diameter,muller2019diameter}, percolation~\cite{serrano2011percolation,fountoulakis2018law}, self-similarity~\cite{serrano2008similarity}, or spectral properties~\cite{kiwi2018spectral} and it has been extended to growing networks~\cite{papadopoulos2012popularity}, as well as to weighted networks~\cite{allard2017geometric}, multilayer networks~\cite{kleineberg2016hidden,Kleineberg2017}, networks with community structure~\cite{zuev2015emergence,garcia-perez:2018aa,muscoloni2018nonuniform} and it is also the basis for defining a renormalization group for complex networks\cite{garcia-perez2018multiscale,Zheng:2021aa}. The analytical tractability of the $\mathbb{S}^1$ model makes it the perfect framework for our work.

In the $\mathbb{S}^1$ model, nodes are assumed to live in a metric similarity space, where similarity refers to all the attributes that control the connectivity in the network, except for the degrees. At the same time, nodes are heterogeneous, with nodes with different levels of popularity coexisting within the same system.  The popularity of a given node is quantified by its hidden degree. In our model, expected degrees can match observed degrees in real networks and we fix the positions of nodes in the metric space so that generated networks can be compared against real networks. This imposes constraints on the connection probability. Specifically, a link between a pair of nodes is created with a probability that resembles a gravity law, increasing with the product of nodes' popularities and decreasing with their distance in the similarity space. We further ask the model to define an ensemble of geometric random graphs with maximum entropy under the constraints of having a fixed expected degree sequence. This determines completely the form of the connection probability depending on the value of one of the model parameters: temperature~\cite{Boguna:2020fj}. Next, we describe the $\mathbb{S}^1$ model in the low and high temperature regimes.

{\subsection*{Low temperature regime}} The $\mathbb{S}^1$ is a model with hidden variables representing the location of the nodes in a similarity space and their popularity within the network. Specifically, each node is assigned a random angular coordinate $\theta_i$ distributed uniformly in $[0,2\pi]$, fixing its position in a circle of radius $R=N/2\pi$. In this way, in the limit $N\gg1$ nodes are distributed in a line according to a Poisson point process of density one with periodic boundary conditions. Each node is also given a hidden degree $\kappa_i$, which corresponds to its ensemble expected degree. In the low temperature regime, each pair of nodes is connected with probability
\begin{equation}
p_{ij}=\frac{1}{1+\left( \frac{x_{ij}}{\hat{\mu} \kappa_i \kappa_j}\right)^\beta},
\label{eq:pij}
\end{equation}
where $x_{ij}=R \Delta \theta_{ij}$ is the distance between nodes $i$ and $j$ along the circle, and $\beta>\beta_c=1$ and $\hat{\mu}$ are model parameters fixing the average clustering coefficient and average degree of the network, respectively. In this representation, the parameter $\beta$ plays the role of the inverse temperature, controlling the level of noise in the system. To see this, the connection probability in Eq.~\eqref{eq:pij} can be rewritten as the Fermi distribution~\cite{Boguna:2020fj}
\begin{equation}
p_{ij}=\frac{1}{e^{\beta(\epsilon_{ij}-\mu)}+1},
\label{eq:fermi}
\end{equation}
where the energy of state $ij$ is 
\begin{equation}
\epsilon_{ij}=\ln{\left[\frac{x_{ij}}{\kappa_i\kappa_j}\right]}
\end{equation}
and where the chemical potential $\mu=\ln{\hat{\mu}}$ fixes the expected number of links, as in the grand canonical ensemble. This result is remarkable as it allows us to map our model to a system of identical particles with Fermi statistics. First, links in our model are unlabeled --and so indistinguishable-- objects. Second, the model generates simple graphs such that only one link can occupy a given state of energy $\epsilon_{ij}$. 
Third, such a state is occupied with the probability given in Eq.~\eqref{eq:fermi}, which is the occupation probability of the Fermi statistics in the grand canonical ensemble. Thus, the $\mathbb{S}^1$ model is equivalent to a system of noninteracting fermions at temperature $T=\frac{1}{\beta}$~\cite{krioukov2010hyperbolic,Boguna:2020fj}. These Fermi-like ``particles'' correspond to the links of the network and live on a discrete phase space defined by the $N(N-1)/2$ pairs among the $N$ nodes of the network. Each such state $ij$ has an associated energy given by $\epsilon_{ij}$, which grows slowly with the distance  between nodes $i$ and $j$ in the metric space.

Despite the fact that links in the model are noninteracting particles, the system undergoes a continuous topological phase transition at a critical temperature $T_c=\beta_c^{-1}=1$, separating a geometric phase, with a finite fraction of triangles attached to nodes induced by the triangle inequality, and a nongeometric phase, where clustering vanishes in the thermodynamic limit~\cite{serrano2008similarity}. We can analyze the nature of the transition by studying the entropy of the ensemble. Given the mapping of the $\mathbb{S}^1$ model to a system of non-interacting fermions in the grand canonical ensemble, we start from the grand canonical partition function
\begin{equation}
    \ln\mathcal{Z}=\sum\limits_{i<j}\ln\left[1+\left(\frac{x_{ij}}{\hat{\mu}\kappa_i\kappa_j}\right)^{-\beta}\right]\label{eq:lnZ},
\end{equation}
where $\hat{\mu}=\exp{\mu}$, with $\mu=\ln{\left( \frac{\beta}{2\pi \langle k \rangle}\sin{\frac{\pi}{\beta}}\right)}$ so that, in the thermodynamic limit, the average degree is set to $\langle k \rangle$~\cite{Boguna:2020fj}. Given the homogeneity and rotational invariance of the distribution of nodes in the similarity space, we can place the $i$'th node on the origin, leading to $N$ identical terms. When the system size is large, we can approximate the sums in Eq. \eqref{eq:lnZ} by integrals. This leads to the following expression
\begin{alignat}{6}
    \ln\mathcal{Z}&=N\iint\mathrm{d}\kappa\mathrm{d}\kappa'\rho(\kappa)\rho(\kappa')\int_0^{\infty}\mathrm{d}x\ln\left[1+\left(\frac{x}{\hat{\mu}\kappa\kappa'}\right)^{-\beta}\right]\nonumber\\[0.3cm]
    &=N\hat{\mu}\langle k \rangle^2\int_0^{\infty}\mathrm{d}t\ln\left[1+t^{-\beta}\right]=N\frac{\hat{\mu}\langle k\rangle^2\pi}{\sin\frac{\pi}{\beta}}.
\end{alignat}
We can then use the above expression to find the grand potential $\Xi=-\beta^{-1}\ln\mathcal{Z}$ and the entropy as $S=\beta^2(\frac{\partial \Xi}{\partial \beta})_{\mu}$
From this, we can find the entropy per link of the system as
\begin{equation}
    \frac{S}{E}=\beta-\pi\cot\frac{\pi}{\beta}\overset{\beta\rightarrow \beta_c^+}{\sim} \frac{1}{\beta-1},
\label{eq:entropy}
\end{equation}
where in the last step $\hat{\mu}$ was plugged in. Note that $E=N\langle k\rangle/2$ is the number of links --and so particles-- in the network. Interestingly, the entropy density is only a function of $\beta$, and so independent of the degree distribution.

\begin{figure}[t]
	\centering
	\includegraphics[width=0.6\linewidth]{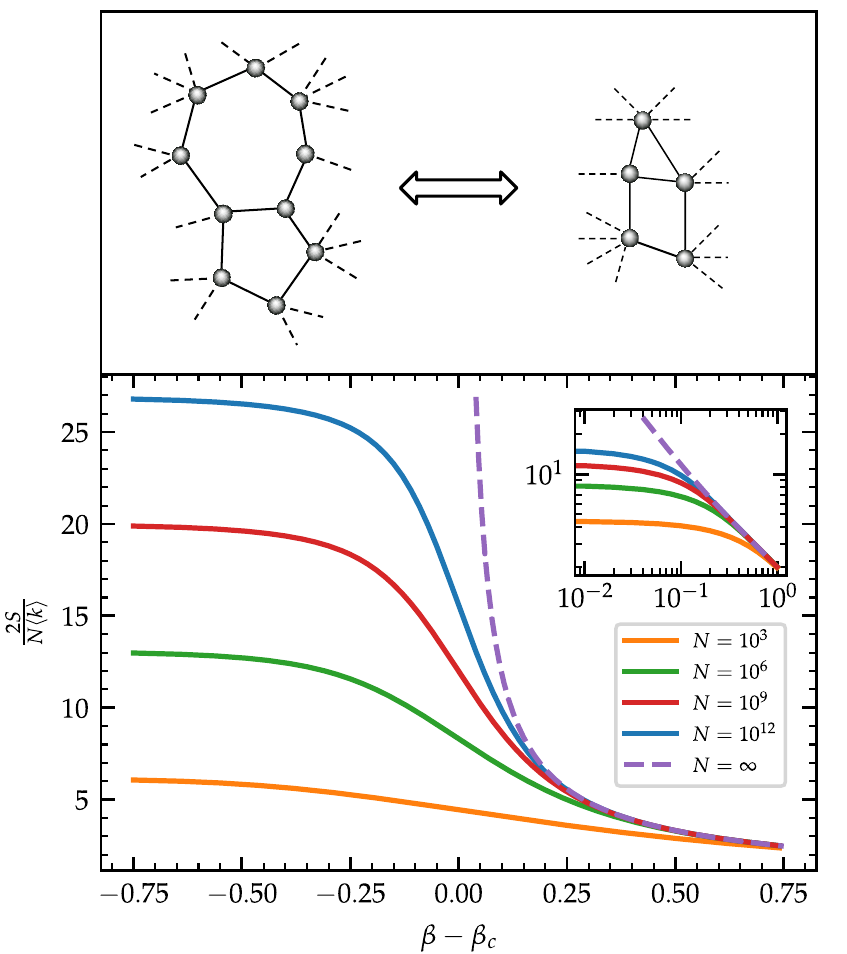}
	\vspace{-0.2cm}\caption{Entropy per link for $\mathbb{S}^1$ geometric networks of different sizes with homogeneous degrees. Different curves are obtained by numerical integration (see SI). The inset shows the same curves in the region $\beta>\beta_c$ in logarithmic scale. The sketch in the top illustrates the different organization of cycles in the two phases, short-range at low temperatures and long-range --of the order of the network diameter-- in the high temperature regime.} 
	\label{fig:1}
\end{figure}

 From Eq.~\eqref{eq:entropy}, we see that the entropy per link diverges at the critical temperature $\beta \rightarrow \beta_c^+=1$. This implies that there is a sudden change in the behavior of the system at the critical point $\beta=\beta_c$, which could indicate the presence of a phase transition. This transition is, however,
 anomalous --at odds with the continuous entropy density usually observed in continuous phase transitions-- and thus cannot be described by Landau's symmetry-breaking theory of continuous phase transitions. Figure~\ref{fig:1} shows a numerical evaluation of the entropy for different system sizes in homogeneous networks confirming the divergence of the entropy per link at the critical temperature as predicted by our analysis. Nevertheless,  as we show in the SI, entropy per link diverges logarithmically with the system size at $\beta=\beta_c$ so that the divergence can only be detected for very large systems.

{\subsection*{High temperature regime}} In the high temperature regime $\beta<\beta_c$ we again fix the angular coordinate and expected degree of the nodes $(\kappa_i,\theta_i)$ so that the degree distribution of the network remains unaltered when temperature is increased beyond the critical point and the model can be directly compared with real networks. Under these constraints, maximizing the entropy of the ensemble leads to the following connection probability~\cite{Boguna:2020fj} 
\begin{equation}
p_{ij}=\frac{1}{1+\frac{x_{ij}^\beta}{\hat{\mu} \kappa_i \kappa_j}},
\label{eq:pij2}
\end{equation}
with $\hat{\mu} \simeq (1-\beta)2^{-\beta}N^{\beta-1}/\langle k \rangle$ for $\beta<1$ and $\hat{\mu} \simeq (2\langle k\rangle\ln N)^{-1}$ when $\beta=1$\footnote{Here we define `$A\simeq B$' as `A is asymptotically equal to B', i.e. that the equality becomes exact as $N\rightarrow\infty$. This in contrast to `$A\sim B$' which means that $A$ and $B$ are asymptotically proportional to one another. }. Notice that this definition of the model converges to the soft configuration model with a given expected degree sequence~\cite{park2004statistical,colomer-de-simon2012clustering,van2018sparse,garlaschelli2018covariance} in the limit of infinite temperature $\beta=0$. As we show in the SI, in this regime long range connections dominate, which causes the entropy density to scale as $\ln{N}$ (see Fig.~\ref{fig:1}) in the whole interval $\beta \in[0,1]$ (and so to diverge in the limit $N\rightarrow \infty$) and the clustering to vanish in the thermodynamic limit.

Notice that the $\mathbb{S}^1$ model is rotationally invariant both above and below the critical temperature, which implies that there is no symmetry breaking at the critical point. In fact, we argue that $\beta_c$ separates two distinct phases with different organization of the cycles, or topological defects, in the network. Indeed, the cycle space of an undirected network with $N$ nodes, $E$ links, and $N_{\text{com}}$ connected components is a vector space of dimension $E-N+N_{\text{com}}$~\cite{gross2018graph}. This dimension is also the number of chordless cycles in the network as they form a complete basis of the cycle space. In complex networks, we are typically interested in connected or quasi-connected networks, with a giant connected component extending almost to the entire network. In the $\mathbb{S}^1$ model this is achieved in the percolated phase when the average degree is sufficiently high, but still in the sparse regime, so that the vast majority of cycles are contained in the giant component. In this case, by changing temperature without changing the degree distribution, the number of nodes, links, and components remain almost invariant and so does the number of chordless cycles. Thus, the two different phases correspond to a different arrangement of the chordless cycles of the network~\footnote{We, however, notice that the preservation of the number of cycles is not a necessary condition for the transition to take place.}, as illustrated in the sketch in Fig.~\ref{fig:1}. This is again similar to the BKT transition since the number of vortices and antivortices is preserved in both phases. 

This difference in arrangement of the cycles is caused by the following process. At low temperatures, the high energy associated to connecting spatially distant points causes the majority of links attached to a given node to be local. This defines the geometric phase at $\beta>\beta_c$ where the triangle inequality plays a critical role in the formation of cycles of finite size. As temperature increases, the number of energetically feasible links connecting very distant pairs of nodes grows, and at $\beta \le \beta_c$ the number of available long range states becomes macroscopic due to the logarithmic dependence of the energy on distance, which causes the entropy per link to be infinite in this regime. This defines a nongeometric phase where links are mainly long ranged and the fraction of finite size cycles vanishes because the triangle inequality stops playing a role. This in turn implies that chordless cycles are necessarily of the order of the network diameter. 

In the geometric phase, there are finite cycles of any order although, as we shown in the SI, the density of triangles is much higher than the density of squares, pentagons, etc. In the nongeometric phase, the cycles are of the order of the network diameter. However, due to the small-world property and finite size effects the diameter of the network can be quite small, so that the distinction between finite cycles of order higher than three and long range cycles can be difficult. Therefore, the average local clustering coefficient --measuring the density of the shortest possible cycles, which are also the most numerous-- is the perfect order parameter to quantify this topological phase transition.

\begin{figure}[t]
	\centering
	\includegraphics[width=0.6\linewidth]{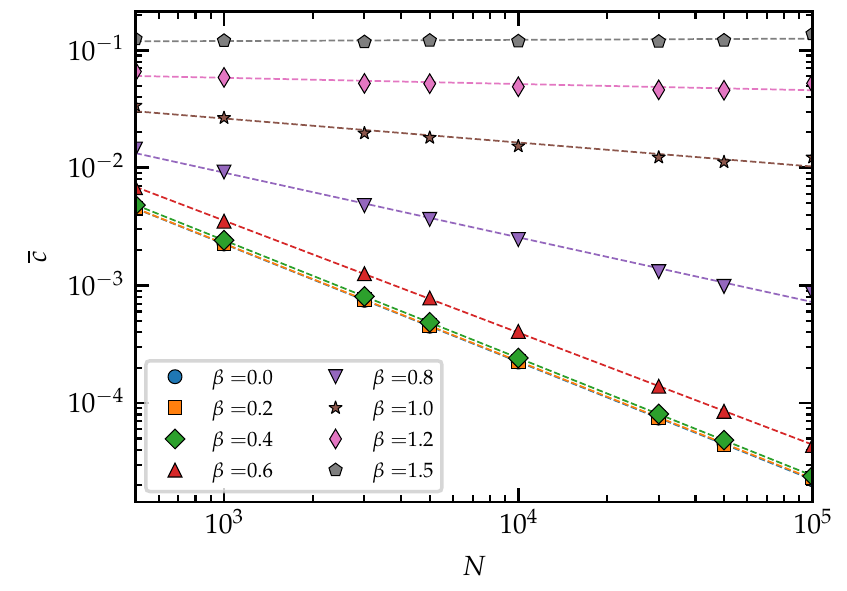}
	\vspace{-0.2cm}\caption{Average local clustering coefficient vs network size for $\mathbb{S}^1$ geometric networks with homogeneous degrees. The networks were generated by applying the DPG technique to a configuration model network with a homogeneous degree sequence $k=4, \forall k$. Dashed lines are power law fits used to estimate the exponent $\sigma$.} 
	\label{fig:0}
\end{figure}
\begin{figure}[t]
	\centering
	\includegraphics[width=0.6\linewidth]{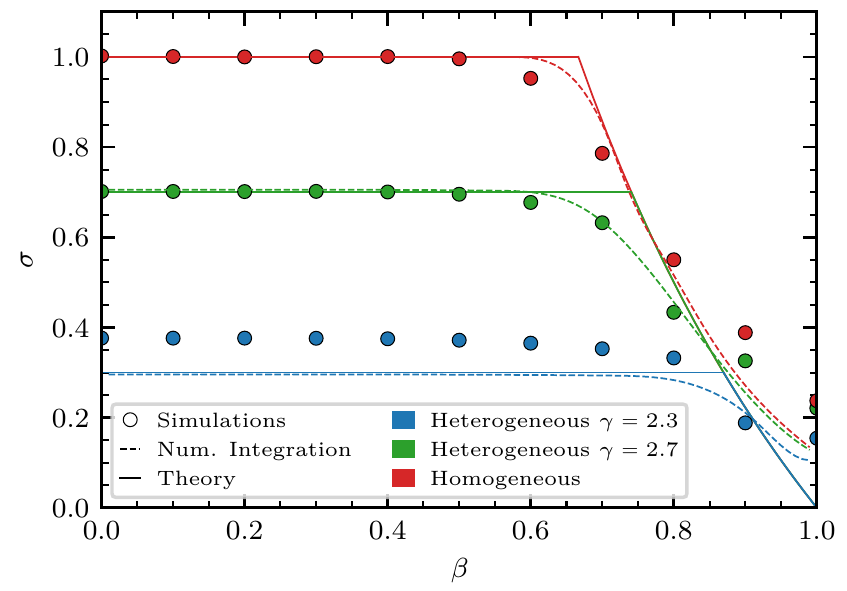}
	\vspace{-0.2cm}\caption{Exponent $\sigma$$(\beta)$, as defined in $\overline{c}(N,\beta)\sim N^{-\sigma(\beta)}$, for $\beta<\beta_c$ evaluated from numerical simulations (colored circles), numerical integration of Eq.~\eqref{eq:clustering} (dashed lines), and theoretical approach Eq.~(\ref{eq:clustering_2},\ref{eq:clustering_3}) (solid lines). Networks are generated with a homogeneous distribution of hidden degrees (red lines and circles) and a power law distribution with exponents $\gamma=2.3$ and $\gamma=2.7$, blue and green lines and circles, respectively.}
	\label{fig:2}
\end{figure}

\subsection*{Finite size scaling of the transition} 
To quantify the behavior of clustering in this transition, we compute the average local clustering coefficient, $\bar{c}$, as the local clustering coefficient averaged over all nodes in a network. The local clustering coefficient for a given node $i$,  with hidden variables $(\kappa_i,\theta_i)$, is defined as the probability that a pair of randomly chosen neighbors are neighbors themselves and, using results from~\cite{boguna2003class}, can be computed as
\begin{equation}
c_i=\frac{\sum_{j \ne i}\sum_{k \ne i}p_{ij}p_{jk}p_{ik}}{\left( \sum_{j\ne i} p_{ij} \right)^2}.
\label{eq:clustering}
\end{equation}
In the SI we derive analytic results for the behavior of the average local clustering coefficient when hidden degrees follow a power law distribution $\rho(\kappa)\sim \kappa^{-\gamma}$ with $2<\gamma<3$ and a cutoff $\kappa<\kappa_c\sim N^{\alpha/2}$. This is done by finding appropriate bounding functions $f(N,\beta)\leq\overline{c}(N,\beta)\leq g(N,\beta)$ that are both asymptotically proportional to $N^{-\sigma(\beta)}$, implying that $\overline{c}\sim N^{-\sigma(\beta)}$ as well. When $\beta>1$, which we call the geometric region, the average local clustering coefficient behaves as~\cite{serrano2008similarity}
\begin{equation}
	\lim_{N\rightarrow\infty}\overline{c}(N,\beta)=Q(\beta),
\end{equation}
for some constant $Q(\beta)$ that depends on $\beta$. Moreover, there exists a constant $Q'$ such that 
\begin{equation}
	\lim_{\beta\rightarrow1^+}\frac{Q(\beta)}{(\beta-1)^2}=Q'\label{eq:clustering_1}
\end{equation}
When $\beta_c'<\beta\leq1$, i.e. in the quasi-geometric region,
\begin{equation}
	\overline{c}(N,\beta)\sim
	\begin{cases}
		(\log N)^{-2}\quad &\text{if } \,\beta=1\\[0.3cm]
		N^{-2(\beta^{-1}-1)}\quad &\text{if } \, \beta_c'<\beta<1\label{eq:clustering_2}
	\end{cases}
\end{equation}
where the vale of $\beta_c'$ depends on the parameter $\alpha$. If $\alpha>1$~\footnote{$\alpha=1$ defines the onset of structural degree-degree correlations~\cite{boguna2004cut}} it is given by $\beta_c'=2/\gamma$ and if $\kappa_c$ grows with $N$ slower than any power law ($\alpha=0$) then $\beta_c'=\frac{2}{3}$. This includes the case of homogeneous degree distributions with $\rho(\kappa)=\delta(\kappa-\langle k \rangle)$ that we study in this paper. Notice that the behavior in a close neighborhood of $\beta_c$ is independent of $\gamma$. The fact that the microscopic details of the model do not affect this scaling behavior points to the universality of our results.\\
Finally, when $\beta<\beta_c'$ (in the non-geometric region), the exact scaling behavior depends on $\alpha$ (see the SI for the general case $\alpha\leq 1$):
\begin{equation}
	\overline{c}(N,\beta)\sim \begin{cases}
		N^{-(\gamma-2)}\log N \quad &\text{if } \, \alpha>1\\[0.3cm]
		N^{-1}\quad &\text{if } \, \alpha=0.\label{eq:clustering_3}
	\end{cases}
\end{equation}
These results are remarkable in many respects. First, clustering undergoes a continuous transition at $\beta_c=1$, attaining a finite value in the geometric phase $\beta>\beta_c$ and becoming zero in the nongeometric phase $\beta<\beta_c$ in the thermodynamic limit. The approach to zero when $\beta \rightarrow \beta_c^+$ is very smooth since both clustering and its first derivative are continuous at the critical point. Second, right at the critical point, clustering decays logarithmically with the system size, and it decays as a power of the system size when $\beta<\beta_c$. This is at odds with traditional continuous phase transitions, where one observes a power law decay at the critical point and an even faster decay in the disordered phase. Third, there is a quasi-geometric region $\beta_c'<\beta<\beta_c$ where clustering decays very slowly, with an exponent that depends on the temperature. Finally, for $\beta<\beta_c'$, we recover the same result as that of the soft configuration model for scale-free degree distributions~\cite{colomer-de-simon2012clustering}. The results in Eqs.~(\ref{eq:clustering_2},\ref{eq:clustering_3}) around the critical point suggest that $N_\text{eff}=\ln{N}$ plays the role of the system size instead of $N$. Indeed, in terms of this effective size, we observe a power law decay at the critical point and a faster decay in the unclustered phase, as expected for a continuous phase transition. Consequently, we expect the finite size scaling ansatz of standard continuous phase transitions to hold with this effective size. We then propose that, in the neighborhood of the critical point, clustering at finite size $N$ can be written as
\begin{equation}
\bar{c}(\beta,N)=\left[\ln{N}\right]^{-\frac{\eta}{\nu}} f\left((\beta-\beta_c)\left[\ln{N}\right]^{\frac{1}{\nu}} \right),
\label{eq:FSS}
\end{equation}
with $\eta=2$, $\nu=1$, and where $f(x)$ is a scaling function that behaves as $f(x) \sim x^\eta$ for $x \rightarrow \infty$.

We test these results with numerical simulations, and by direct numerical integration of Eq.~\eqref{eq:clustering} using Eq.~(\ref{eq:pij}) for $\beta>\beta_c$ and Eq.~(\ref{eq:pij2}) for $\beta \leq \beta_c$, see SI. Simulations are performed with the degree-preserving geometric (DPG) Metropolis-Hastings algorithm introduced in~\cite{Starnini_2019}, that allows us to explore different values of $\beta$ while preserving exactly the degree sequence. Given a network, the algorithm selects at random a pair of links connecting nodes $i,j$ and $l,m$ and swaps them (avoiding multiple links and self-connections) with a probability given by
\begin{equation}
p_{swap}=\min{\left[1,\left( \frac{\Delta \theta_{ij}\Delta \theta_{lm}}{\Delta \theta_{il}\Delta \theta_{jm}}\right)^{\beta}\right]},
\label{DP:metropolis}
\end{equation}
where $\Delta \theta$ is the angular separation between the corresponding pair of nodes. This algorithm maximizes the likelihood that the network is $\mathbb{S}^1$ geometric while preserving the degree sequence and the set of angular coordinates, and it does so independently of whether the system is above or below the critical temperature.  Notice that the continuity of Eq.~\eqref{DP:metropolis} as a function of $\beta$ makes it evident that, even if the connection probability takes a different functional form above and below the critical point, the model is the same.

\begin{figure}[t]
	\centering
	\includegraphics[width=0.6\linewidth]{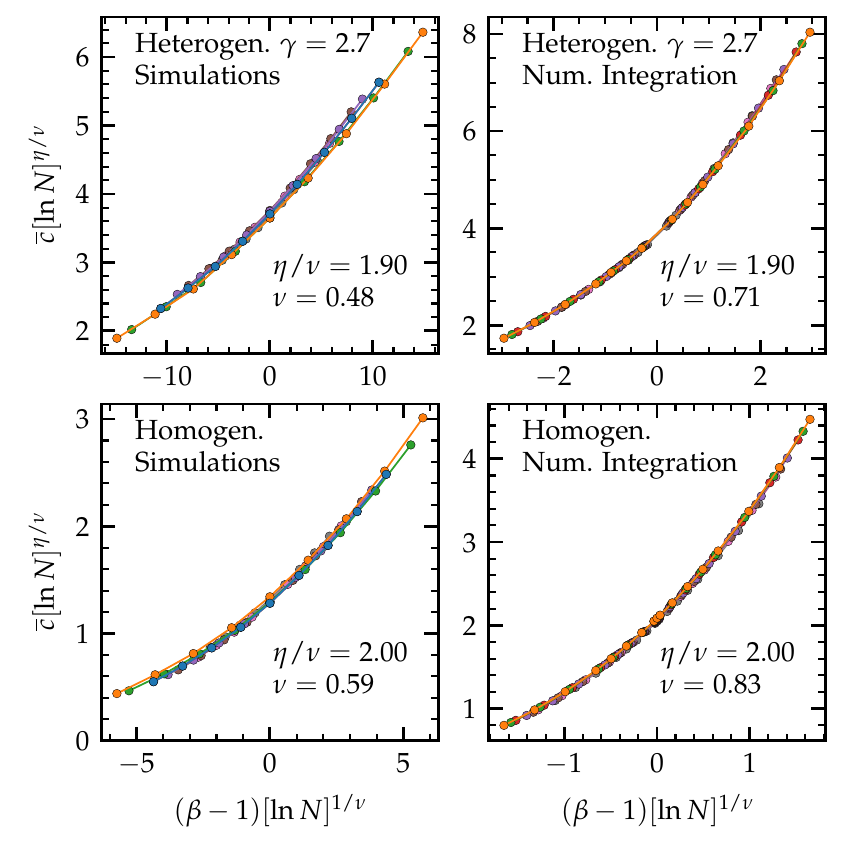}
	\vspace{-0.2cm}\caption{Finite size scaling Eq.~\eqref{eq:FSS} for heterogeneous networks with $\gamma=2.7$ (top row) and homogeneous networks (bottom row). Left column correspond to numerical simulations with sizes in the range $N\in(5\times10^2,10^5)$, whereas the right column is obtained from numerical integration of Eq.~\eqref{eq:clustering} with sizes in the range $N \in (5\times10^5,10^8)$. Different colors correspond to the different system sizes used.}
	\label{fig:3}
\end{figure}

Figure~\ref{fig:0} shows the behavior of the average local clustering coefficient as a function of the number of nodes for homogeneous $\mathbb{S}^1$ networks with different values of $\beta$, showing a clear power law dependence $N^{-\sigma(\beta)}$ in the nongeometric phase $\beta<\beta_c$, with an exponent that varies with $\beta$ as predicted by our analysis. These results are used to measure the exponent $\sigma(\beta)$ as a function of the inverse temperature $\beta$, which in Fig.~\ref{fig:2} are compared with the theoretical value given by Eq.~(\ref{eq:clustering_2},\ref{eq:clustering_3}). The agreement is in general very good, although it gets worse for values of $\beta$ very close to $\beta_c$ and for very heterogeneous networks. This discrepancy is expected due to the slow approach to the thermodynamic limit in the nongeometric phase, which suggests that the range of our numerical simulations, $N\in[5\times 10^2,10^5]$, is too limited. To test for this possibility, we solve numerically Eq.~\eqref{eq:clustering} for sizes in the range $N\in[5\times10^5,10^8]$ and measure numerically the exponent $\sigma(\beta)$. In this case, the agreement is also very good for heterogeneous networks. The remaining discrepancy when $\beta \approx \beta_c$ is again expected since, as shown in Eq.~\eqref{eq:clustering_2}, right at the critical point clustering decays logarithmically rather than as a power law. Finally, Fig.~\ref{fig:3} shows the finite size scaling Eq.~\eqref{eq:FSS} both for the numerical simulations and numerical integration of Eq~\eqref{eq:clustering}. In both cases, we find a very good collapse with exponent $\eta/\nu \approx 2$ in all cases. The exponent $\nu$, however, departs from the theoretical value $\nu=1$ in numerical simulations due to their small sizes but improves significantly with numerical integration for bigger sizes. We then expect Eq.~\eqref{eq:FSS} to hold, albeit for very large system sizes.

The slow decay of clustering in the nongeometric phase implies that some real networks with significant levels of clustering may be better described using the $\mathbb{S}^1$ model with temperatures in the quasi-geometric regime $\beta<\beta_c$. Given a real network, the DPG algorithm can be used to find its value of $\beta$. To do so, nodes in the real network are given random angular coordinates in $(0,2\pi)$. Then the DPG algorithm is applied, increasing progressively the value of $\beta$ until the average local clustering coefficient of the randomized network matches the one measured in the real network. Many real networks have very high levels of clustering and lead to values of $\beta>\beta_c$. However, there are notable cases with values of $\beta$ below the critical point. As an example, in the SI table
we show values of $\beta$ obtained for several real networks with values below or slightly above $\beta_c$. In fact, some of them are found to be very close to the critical point, like protein-protein interaction networks of specific human tissues\cite{Chang2014}, with $\beta \approx 1$, or the genetic interaction network of the Drosophila Melanogaster\cite{Stark2006}, $\beta \approx 1.1$.

\section*{Conclusion}
Our results in this paper show that the dependence of the phase space on an underlying geometry in networks where edges are considered noninteracting particles leads to an anomalous phase transition between different topological orders. Despite particles being noninteracting, the set of states that they can occupy are correlated by the triangle inequality in the underlying metric space. This correlation induces an effective interaction between particles, ultimately leading to a clustered phase at low temperatures. Interestingly, the logarithmic dependence of the state-energy with the metric distance results in the divergence of the entropy at a finite temperature $\beta_c$ and, thus, to a different ordering of cycles below $\beta_c$, where clustering vanishes in the thermodynamic limit. The finite size behavior of the transition is anomalous, with $\ln{N}$ and not $N$ playing the role of the system size. This slow approach to the thermodynamic limit is relevant for real networks in the quasi-geometric phase $\beta_c'<\beta<1$
, for which high levels of clustering can still be observed. All together, our results describe an anomalous topological phase transition that cannot be described by the classic Landau theory but that, nevertheless, differs from other topological phase transitions, such as the BKT transition, in the behavior of thermodynamic properties.

%
\section*{Acknowledgments}
We acknowledge support from: Agencia estatal de investigaci\'on project number PID2019-106290GB-\linebreak C22/AEI/10.13039/501100011033; and Generalitat de Catalunya grant number 2017SGR1064. M. B. acknowledges the ICREA Academia award, funded by the Generalitat de Catalunya. J.~vd~K. acknowledges support from the Secretaria d'Universitats i Recerca de la Generalitat de Catalunya i del Fons Social Europeu.

\bibliographystyle{nature}
\bibliography{geometry3}

\section*{Authors contributions} All authors participated in the design and implementation of the research, analysis of results, and in the writing of the paper. 

\section*{Competing interests} The authors declare no competing interests.

\end{document}



\baselineskip24pt


\maketitle 

\tableofcontents

\section{Analytics}
In the following section we will first present some preliminaries about the $\mathbb{S}^1$-model. We will do so from the point of view of network theory, focussing on connection probabilities and hidden variables of nodes. This section functions as a summary of known results about the $\mathbb{S}^1$-model. For further reading about this model and its alternative formulation, the $\mathbb{H}^2$-model, we kindly direct the reader to~\cite{serrano_boguna_2022}. We will then look at the model from another direction, focussing on the fact that the network can be represented as a gas of fermions (links), where the nodes of the network define the available states. This will allow us to fully determine the thermodynamics of the model, generalizing what was already done in the main text of the paper. We investigate the surprising result further by showing that many can be recovered by a highly simplified toy-model. We then revert back to the network point of view to find the finite size scaling behaviour of the clustering coefficient in different cases. This is done by looking at $N\gg1$ but finite, taking only the dominant contributions into account. We distinguish between the cases $\beta<\beta_c$ and $\beta=\beta_c$ as these show different behaviour. Finally we will study how the clustering coefficient in the thermodynamic limit ($N\rightarrow\infty$) approaches zero in the limit $\beta\rightarrow\beta_c^+$. Notation-wise we choose to use $a\simeq b$ to refer to `$a$ is asymptotically equivalent to $b$' and $\sim$ to `$a$ is asymptotically proportional to $b$'.
\subsection{Preliminaries}
The average clustering coefficient for a node with hidden degree $\kappa$ and angular position $\theta$ is defined in the $\mathbb{S}^1$~\cite{boguna2003class} model as
\begin{equation}
    \overline{c}(\kappa,\theta)=\left(\frac{N}{\bar{k}(\kappa,\theta)}\right)^2\iiiint \mathrm{d} \kappa'\mathrm{d}\kappa''\mathrm{d}\theta'\mathrm{d}\theta'' \rho(\kappa')\rho(\kappa'')p(\kappa',\kappa'',\theta',\theta'')p(\kappa,\kappa'',\theta,\theta'')p(\kappa,\kappa',\theta,\theta')\label{eq:clustering},
\end{equation}
in a network with system size $N$. Here the function $\bar{k}(\kappa,\theta)$ is the average degree of a node with hidden coordinates $(\kappa,\theta$), $\rho(\kappa)$ is the hidden degree density and $p(\kappa,\kappa',\theta,\theta'')$ is the connection probability between two nodes with hidden coordinates $(\kappa,\theta)$ and $(\kappa',\theta')$. The exact form of these functions will be discussed in the following. Note that as the model has rotation symmetry, one only needs to investigate the node at angular coordinate $\theta=0$. The average clustering coefficient can be computed from $\overline{c}(\kappa)$ in the following manner
\begin{equation}
    \overline{c}=\int\mathrm{d}\kappa'\rho(\kappa')\overline{c}(\kappa').
\end{equation}
However, as $\overline{c}(\kappa)$ is a bounded monotonically decreasing function, it suffices to find the scaling of $\overline{c}(\kappa)$ for small $\kappa$~\cite{colomer-de-simon2012clustering}. In Eq.~\eqref{eq:clustering}, $\rho(\kappa)$ defines the distribution of the hidden degrees. In the following, we always apply a power law distribution. As we are interested in finite-sized scale-free networks, we choose the following distribution
\begin{equation}
	\rho(\kappa)=\begin{cases}
		\frac{(\gamma-1)\kappa_0^{\gamma-1}}{1-\left(\frac{\kappa_c}{\kappa_0}\right)^{1-\gamma}}\kappa^{-\gamma}\,& \text{if }\quad \kappa_0\leq\kappa\leq\kappa_c\\
		0& \text{else}
	\end{cases}
\label{eq:hiddendegreedensity}
\end{equation}
Note that from this expression also the homogeneous distribution can be reached. This can be done in two different ways. The first is by letting $\kappa_c\rightarrow\kappa_0$. Note that then $\rho(\kappa)\rightarrow\infty$ if $\kappa_0\leq\kappa\leq\kappa_c$, but that this region also goes to zero width. Thus, we end up with a delta distribution (note that the integral of $\rho(\kappa)$ always gives $1$, irrespective of $\kappa_c$), exactly what we want for a homogeneous distribution. We can then set $\kappa_0=\langle k\rangle$ to obtain the correct average degree. One can make similar arguments by leaving $\kappa_c>\kappa_0$ and $\gamma\rightarrow\infty$. In this case, one again ends up with the same distribution. We choose to not specify the specific form of the cut-offs just yet. We just demand that $\kappa_0$ is such that the correct average degree is obtained and that, to lowest order, it does not depend on the system size. The average degree of nodes with hidden variable $\kappa$ and angular position $\theta$ is defined as
\begin{equation}
    \bar{k}(\kappa,\theta)=N\iint\mathrm{d}  \kappa'\mathrm{d} \theta' \rho(\kappa')p(\kappa,\kappa',\theta,\theta').\label{eq:av.degree}
\end{equation}
The function $p$ describes the probability of two nodes in the network being connected and is given by the Fermi-Dirac distribution. In Ref.~\cite{Boguna:2020fj} it is noted that for $\beta>\beta_c$, one can define the connection probability in the thermodynamic limit, given in terms of the spatial coordinates $r$, in our 1D case the coordinates on a infinite line:
\begin{equation}
	p(\kappa',\kappa'',r',r'')=\dfrac{1}{1+\left(\dfrac{|r'-r''|}{\hat{\mu}\kappa'\kappa''}\right)^\beta},\label{eq:connectionprobabilityThermolimit}
\end{equation}
Here $\hat{\mu}=\exp \mu$ where $\mu$ is the chemical potential which fixes the average degree of the network. We come back to this shortly. As was noted in the main text, the relation between the coordinate $\theta$ on a circle with a finite radius and the coordinate on the real line $r$ is $r=\frac{N\theta}{2\pi}$. So for finite sizes this becomes
\begin{equation}
	p(\kappa',\kappa'',\theta',\theta'')=\frac{1}{1+\left(\dfrac{N\Delta\theta}{2\pi\hat{\mu}\kappa'\kappa''}\right)^{\beta}}\label{eq:Fbetabig}.
\end{equation}
Here $\Delta\theta=\pi-|\pi-|\theta'-\theta''||$. To find the value of $\hat{\mu}$ we demand that the average degree remains constant:
\begin{alignat}{6}
	\langle k \rangle &= \frac{N}{\pi}\iint\mathrm{d}\kappa'\mathrm{d}\kappa''\rho(\kappa')\rho(\kappa'')\int^\pi_0\mathrm{d}\theta'\frac{1}{1+\left(\frac{N\theta'}{2\pi\hat{\mu}\kappa'\kappa''}\right)^\beta} \label{eq:fixingmu}\\
	&=N\iint\mathrm{d}\kappa'\mathrm{d}\kappa''\rho(\kappa')\rho(\kappa'')\,{}_2F_1\left[
	\begin{array}{c}
		1,1/\beta\\
		1+1/\beta\end{array};-\left(\frac{N}{2\hat{\mu}\kappa'\kappa''}\right)^\beta\right]\label{eq:muexact}\\
	&\simeq\frac{2\pi\hat{\mu}\langle k\rangle^2}{\beta\sin\left(\pi/\beta\right)}+(2\hat{\mu})^\beta\frac{N^{1-\beta}}{1-\beta}\kappa_0^{2\beta}\left(\frac{\gamma-1}{\gamma-\beta-1}\right)^{2}\label{eq:muapprox}
\end{alignat}
Here, $_2F_1\left[
\begin{array}{c}
	a,b\\
	c\end{array};z\right]$ is the ordinary hypergeometric function~\footnote{Note that we use a slightly different form than the standard $_2F_1[a,b;c;z]$ for aesthetic purposes.}. Which one of these terms is dominant depends on $\beta$. If $\beta>1$, the first term is more important and so we can isolate $\hat{\mu}$ to obtain
\begin{equation}
	\hat{\mu}\simeq\frac{\beta\sin(\pi/\beta)}{2\pi\langle k\rangle}.\label{eq:mulowtemp}
\end{equation}
If $\beta<1$, the second term dominates and we obtain
\begin{equation}
	\hat{\mu}\simeq\frac{1}{2\kappa_0^2}(1-\beta)^{1/\beta}\langle k \rangle^{1/\beta}N^{1-1/\beta}\left(\frac{\gamma-\beta-1}{\gamma-1}\right)^{2/\beta}.\label{eq:muhightempER}
\end{equation}
However, as explained in the main text, using connection probability \eqref{eq:Fbetabig} also when $\beta<1$ leads to an ever more homogeneous network. If instead we want to preserve the degree sequence also below the critical $\beta$ we need to redefine the connection probability in this regime:
\begin{equation}
	p(\kappa',\kappa'',\theta',\theta'')=\frac{1}{1+\dfrac{(N\Delta\theta)^\beta}{(2\pi)^\beta\hat{\mu}\kappa'\kappa''}}\label{eq:Fbetasmall}.
\end{equation}
If we use this connection probability instead in Eq. \eqref{eq:fixingmu} we obtain for $\hat{\mu}$ when $\beta<1$
\begin{equation}
	\hat{\mu}\simeq\frac{(1-\beta)}{2^\beta\langle k\rangle N^{1-\beta}}.\label{eq:muhightemp}
\end{equation}
It can be shown that higher order terms become relevant when $\beta\rightarrow 1$. Thus, when $\beta=1$ the form of the dominant contribution to chemical potential will change. Note that in this case both connection probabilities are equivalent. To fix the average degree we write
\begin{alignat}{6}
	\langle k \rangle &= \frac{N}{\pi}\iint\mathrm{d}\kappa'\mathrm{d}\kappa''\rho(\kappa')\rho(\kappa'')\int^\pi_0\mathrm{d}\theta'\frac{1}{1+\left(\frac{N\theta'}{2\pi\hat{\mu}\kappa'\kappa''}\right)} \nonumber\\
	&= N\iint\mathrm{d}\kappa'\mathrm{d}\kappa''\rho(\kappa')\rho(\kappa''){}_2F_1\left[
	\begin{array}{c}
		1,1\\
		2\end{array};-\frac{N}{2\hat{\mu}\kappa'\kappa''}\right]\simeq 2\langle k\rangle^2\hat{\mu}\ln(N)
\end{alignat}
This then leads to 
\begin{equation}
	\hat{\mu}\simeq\Big(2\langle k\rangle\ln(N)\Big)^{-1}\label{eq:mub1}.
\end{equation}
Having derived the expressions above, we can also determine $\overline{k}(\kappa,\theta)$. For large $N$, Eq.~\eqref{eq:av.degree} evaluates to $\overline{k}(\theta,\kappa)\simeq\mathcal{C}\hat{\mu}\langle \kappa\rangle\kappa$, where $\mathcal{C}$ is some constant that depends on $\beta$. Thus, we note that the expected degree is proportional to the hidden degree. Now, integrating over $\kappa$ we obtain $\langle k\rangle \simeq \mathcal{C}\hat{\mu}\langle \kappa\rangle^2$, where above we have defined the various $\hat{\mu}$'s s.t. $\langle k\rangle=\langle\kappa\rangle$. This then implies that $\overline{k}(\theta,\kappa)\simeq\kappa$, i.e. that the hidden degree exactly represents the expected degree of a node.

\subsection{The network as a gas of fermions}	
We will now look at the network in a different picture, using the fact that, as explained in the main text, the edges can be seen as fermions, with occupation numbers given by $(1+\exp{(\beta(\epsilon-\mu))})^{-1}$.
\subsubsection{The density of states}
We start from the most general form of the connection probability $$p=\frac{1}{1+e^{\beta(\epsilon-\mu)}}.$$ Now, if we want the connection probability to have the form as given in Eq. \eqref{eq:Fbetabig}, where $\hat{\mu}=\exp(\mu)$, we must define the energy per link/particle as 
\begin{equation}
	\epsilon(\theta',\theta'',\kappa',\kappa'') = \ln\left(\frac{N(\pi-|\pi-|\theta'-\theta''||)}{2\pi\kappa'\kappa''}\right)\label{eq:el}.
\end{equation}
As was mentioned above as well as in the main text, we must change the form of the connection probability for $\beta<\beta_c$ in order to have a degree distribution independent of temperature. The form we then use is  that given in Eq. \eqref{eq:Fbetasmall}, where $\hat{\mu}=\exp{\beta\mu}$, which leads to the following energy per particle.  
\begin{equation}
	\epsilon(\theta',\theta'',\kappa',\kappa'') = \ln\left(\frac{N(\pi-|\pi-|\theta'-\theta''||)}{2\pi(\kappa'\kappa'')^{1/\beta}}\right)\label{eq:es}.
\end{equation}
Note that, from a standard statistical physics perspective having the energy levels depend on temperature explicitly is unusual. In fact, we will see that we need to be very careful when deriving the thermodynamic properties. However, we will also show that, from a network perspective, the results we obtain are completely valid. \\

With the two expressions for the energy per particles we can then derive the density of states as follows:
\begin{equation}
	\rho(\epsilon)=\int_0^{2\pi}\mathrm{d}\theta'\rho(\theta')\int_0^{2\pi}\mathrm{d}\theta''\rho(\theta'')\int^{\infty}_{\kappa_0}\mathrm{d}\kappa'\rho(\kappa')\int^{\infty}_{\kappa_0}\mathrm{d}\kappa''\rho(\kappa'')\delta(\epsilon-\epsilon(\theta',\theta'',\kappa',\kappa''))
\end{equation}
This leads to two distinct density of states. The first, using Eq. \eqref{eq:el}, is
\begin{equation}
	\rho(\epsilon)=2\left(\frac{\gamma-1}{2-\gamma}\right)^{2}\kappa_0^{4}e^{\epsilon+\epsilon_{\text{max}}}\Theta(\epsilon_{\text{max}}-\epsilon)\left[1+e^{(2-\gamma)(\epsilon_{\text{max}}-\epsilon)}((2-\gamma)(\epsilon_{\text{max}}-\epsilon)-1)\right]\label{eq:rhoel},
\end{equation}
with $\epsilon_{\text{max}}=\ln\left(\frac{N}{2\kappa_0^2}\right)$ (connecting two points on opposite sides of the $\mathbb{S}^1$ manifold with both points having the minimal expected degree)	and the second, using Eq. \eqref{eq:es}, is
\begin{equation}
	\rho(\epsilon)=2\left(\frac{\gamma-1}{1+1/\beta-\gamma}\right)^{2}\kappa_0^{4/\beta}e^{\epsilon+\epsilon_{\text{max}}}\Theta(\epsilon_{\text{max}}-\epsilon)\left[1+e^{(\beta+1-\beta\gamma)(\epsilon_{\text{max}}-\epsilon)}((\beta+1-\beta\gamma)(\epsilon_{\text{max}}-\epsilon)-1)\right]\label{eq:rhoes},
\end{equation}
with $\epsilon_{\text{max}}=\ln\left(\frac{N}{2\kappa_0^{2/\beta}}\right)$. Note that at $\beta=1$ these two are the same.
The general form is thus
\begin{equation}
	\rho(\epsilon)=ae^{\epsilon+\epsilon_{\text{max}}}\Theta(\epsilon_{\text{max}}-\epsilon)\left[1+e^{b(\epsilon_{\text{max}}-\epsilon)}(b(\epsilon_{\text{max}}-\epsilon)-1)\right]\label{eq:dosgen}.
\end{equation}
\subsubsection{Chemical Potential}
With this we can calculate the chemical potential. In order to do so we study the average amount of links
\begin{alignat}{6}
	\langle E\rangle&=\int_{-\infty}^{\epsilon_{\text{max}}}\mathrm{d}\epsilon\frac{\rho(\epsilon)}{1+e^{\beta(\epsilon-\mu)}}&&\nonumber\\[0.5mm]
	&=ae^{2\epsilon_{\text{max}}}\bigg(\frac{b^2}{(b-1)^2}+e^{\beta(\epsilon_{\text{max}}-\mu)}\bigg(&&-\frac{b\Phi\big[-e^{\beta(\epsilon_{\text{max}}-\mu)},2,\frac{1-b+\beta}{\beta}\big]}{\beta^2}-\frac{{}_2F_1\left[\begin{array}{c}
			1,1+\frac{1}{\beta}\\[0.5mm]
			2+\frac{1}{\beta}
		\end{array};-e^{\beta(\epsilon_{\text{max}}-\mu)}\right]}{1+\beta}\nonumber\\[0.5mm]
	&&&+\frac{{}_2F_1\left[\begin{array}{c}
			1,1+\frac{1-b}{\beta}\\[0.5mm]
			2+\frac{1-b}{\beta}
		\end{array};-e^{\beta(\epsilon_{\text{max}}-\mu)}\right]}{1+\beta-b}\bigg)\bigg).
\end{alignat}
Here, $\Phi[z,a,b]$ is the Lerch zeta function. If we now assume $e^{\beta(\epsilon_{\text{max}}-\mu)}\gg 1$, we can approximate this as
\begin{alignat}{6}
	\langle E\rangle&\simeq ae^{(2+\beta)\epsilon_{\text{max}}-\beta\mu}\bigg\{\frac{1}{\beta}\pi\csc\left(\frac{\pi}{\beta}\right)e^{-(1+\beta)(\epsilon_{\text{max}}-\mu)}+\frac{b^2}{(1-\beta)(b+\beta-1)^2}e^{-2\beta(\epsilon_{\text{max}}-\mu)}\bigg\}.
\end{alignat}
We know that $\epsilon_{\text{max}}\sim \ln N$ so $\mu\simeq c\ln N$ where $c<1$. It can then be shown that for all $c$ the dominant contributions are
\begin{equation}
	\langle E\rangle\simeq
	\begin{cases}
		\frac{a\pi}{\beta}e^{\epsilon_{\text{max}}+\mu}\csc\left(\frac{\pi}{\beta}\right)&\quad\text{if}\quad\beta>1\\[0.5cm]
		a\epsilon_{\text{max}}e^{\epsilon_{\text{max}}+\mu}&\quad\text{if}\quad\beta=1\\[0.5cm]
		\frac{ab^2}{(1-\beta)(b+\beta-1)^2}e^{(2-\beta)\epsilon_{\text{max}}+\mu\beta}&\quad\text{if}\quad\beta<1\label{eq:Ns}
	\end{cases}
\end{equation}
If we take $\langle E\rangle=N\langle k\rangle/2$ (sparse network) we obtain
\begin{alignat}{6}
	\mu\simeq
	\begin{cases}
		\ln\left(\frac{\beta\sin\left(\frac{\pi}{\beta}\right)}{2\pi\langle k\rangle}\right)&\text{ if }\beta>1\\[0.5cm]
		\frac{1}{2\langle k\rangle \ln N}&\text{ if }\beta=1 \\[0.5cm]
		\frac{1}{\beta}\ln\left(\frac{N^{\beta-1}(1-\beta)}{2^{\beta}\langle k\rangle}\right)&\text{ if }\beta<1
	\end{cases}
\end{alignat}
Note that in all  these cases $e^{\beta(\epsilon_{\text{max}}-\mu)}\gg 1$ and that these are exactly the same results as we found before.

\subsubsection{Thermodynamics}
With this we can now study the grand potential. 
\begin{alignat}{6}
	\Xi&=-\frac{1}{\beta}\int_{-\infty}^{\epsilon_{\text{max}}}\mathrm{d}\epsilon\rho(\epsilon)\nonumber\ln\left(1+e^{-\beta(\epsilon-\mu)}\right)\nonumber\\[0.5mm]
	&=-\frac{a}{\beta}e^{2\epsilon_{\text{max}}}\bigg\{\frac{b}{\beta(1-b)}\Phi\bigg[-e^{\beta(\epsilon_{\text{max}}-\mu)},2,\frac{1-b}{\beta}\bigg]+(-1)^{-1/\beta}e^{-(\epsilon_{\text{max}}-\mu)}B_{-e^{\beta(\epsilon_{\text{max}}-\mu)}}\bigg[1+1/\beta,0\bigg]\nonumber\\[0.5mm]
	&\frac{1-2b+(b-1)b\epsilon_{\text{max}}}{1-b+\beta}\frac{\beta}{(1-b)^2}{}_2F_1\left[\begin{array}{c}
		1,1+\frac{1-b}{\beta}\\[0.5mm]
		2+\frac{1-b}{\beta}
	\end{array};-e^{\beta(\epsilon_{\text{max}}-\mu)}\right]e^{\beta(\epsilon_{\text{max}}-\mu)}\nonumber\\[0.5mm]
	&+\frac{\beta b}{(b-1)^3}(1-\epsilon_{\text{max}}+b(-3+b+\epsilon_{\text{max}}))+\frac{b^2}{(1-b)^2}\ln\left(1+e^{-\beta(\epsilon_{\text{max}}-\mu)}\right)\nonumber\\[0.5mm]
	&-\frac{b\beta}{(1-b)^2}\epsilon_{\text{max}}{}_2F_1\left[\begin{array}{c}
		1,\frac{1-b}{\beta}\\[0.5mm]
		1+\frac{1-b}{\beta}
	\end{array};-e^{\beta(\epsilon_{\text{max}}-\mu)}\right]e^{\beta(\epsilon_{\text{max}}-\mu)}\bigg\},
\end{alignat}
where $B_z[a,b]$ is the incomplete beta function. Again assuming that $e^{\beta(\epsilon_{\text{max}}-\mu)}\gg 1$ and $b<1$ we get the following dominant terms, after having divided out $\langle E\rangle$
\begin{alignat}{6}
	\frac{\Xi}{E}\simeq\begin{cases}
		-1&\quad\text{if}\quad\beta>1\\[0.5cm]
		-1&\quad\text{if}\quad\beta=1\\[0.5cm]
		-\frac{1}{\beta}&\quad\text{if}\quad\beta<1
	\end{cases}
\end{alignat}
Normally with this we have enough to calculate the entropy. However, we need to be careful when using a temperature dependent density of states. Let us check if $S=\beta^2\left(\frac{\partial \Xi}{\partial\beta}\right)_\mu$ still holds.
\begin{alignat}{6}
	\beta^2 \left(\frac{\partial \Xi}{\partial\beta}\right)_\mu &= \underbrace{-\beta \rho(\epsilon_{\text{max}})\frac{\partial \epsilon_{\text{max}}}{\partial \beta}\ln\left(1+e^{-\beta(\epsilon_{\text{max}}-\mu)}\right)-\beta\int^{\epsilon_{\text{max}}}_{-\infty}\mathrm{d}\epsilon\frac{\partial\rho(\epsilon)}{\partial\beta}\ln\left(1+e^{-\beta(\epsilon_{\text{max}}-\mu)}\right)}_{\Delta}\nonumber\\[0.5mm]
	&+\underbrace{\int^{\epsilon_{\text{max}}}_{-\infty}\mathrm{d}\epsilon\rho(\epsilon)\bigg(\ln\left(1+e^{-\beta(\epsilon_{\text{max}}-\mu)}\right)+\beta(\epsilon-\mu)\frac{1}{1+\epsilon^{\beta(\epsilon-\mu)}}\bigg)}_{-\beta\Xi+\beta(\langle U\rangle-\mu\langle E\rangle)}\nonumber\\[0.5mm]
	&=\Delta+\int^{\epsilon_{\text{max}}}_{-\infty}\mathrm{d}\epsilon\rho(\epsilon)\bigg(\frac{\ln\left(1+e^{\beta(\epsilon_{\text{max}}-\mu)}\right)}{1+\epsilon^{\beta(\epsilon-\mu)}}+\frac{\ln\left(1+e^{-\beta(\epsilon_{\text{max}}-\mu)}\right)}{1+\epsilon^{-\beta(\epsilon-\mu)}}\bigg)\nonumber\\[0.5mm]
	&=\Delta-\int^{\epsilon_{\text{max}}}_{-\infty}\mathrm{d}\epsilon\rho(\epsilon)\underbrace{\bigg(\frac{1}{1+\epsilon^{\beta(\epsilon-\mu)}}}_{p(\epsilon)}\ln\left(\frac{1}{1+e^{\beta(\epsilon_{\text{max}}-\mu)}}\right)+\left(1-\frac{1}{1+\epsilon^{\beta(\epsilon-\mu)}}\right)\ln\left(1-\frac{1}{1+e^{\beta(\epsilon_{\text{max}}-\mu)}}\right)\bigg)\nonumber\\[0.5mm]
	&= \Delta-\int^{\epsilon_{\text{max}}}_{-\infty}\mathrm{d}\epsilon\rho(\epsilon)\bigg(p(\epsilon)\ln(p(\epsilon))+(1-p(\epsilon))\ln(1-p(\epsilon))\bigg)= \Delta+S
\end{alignat}
In the last step we recognize the entropy of a graphon gas \cite{Voitalov2020}. So, indeed, in the case that $\rho(\epsilon)$ or $\epsilon_{\text{max}}$ depends on the temperature, we get extra terms ($S=\beta^2 \left(\frac{\partial \Xi}{\partial\beta}\right)_\mu-\Delta$). These terms, at least in the general case, are not trivial to evaluate. However, we also note that $S=\beta(\langle U\rangle-\Xi-\mu\langle E\rangle)$ remains valid in all cases. We will therefore approach S in this way. Thus, the final thing we need to do is find an expression for the average energy.  
\begin{alignat}{6}
	\langle U\rangle&=\int_{-\infty}^{\epsilon_{\text{max}}}\mathrm{d}\epsilon\frac{\epsilon\rho(\epsilon)}{1+e^{\beta(\epsilon-\mu)}}\nonumber\\[0.5mm]
	&=ae^{2\epsilon_{\text{max}}}\Bigg\{{}_2F_1\left[\begin{array}{c}
		1,\frac{1-b}{\beta}\\[0.5mm]
		1+\frac{1-b}{\beta}
	\end{array};-e^{\beta(\epsilon_{\text{max}}-\mu)}\right]\epsilon_{\text{max}}+\frac{1}{b-1}{}_2F_1\left[\begin{array}{c}
	1,\frac{1-b}{\beta}\\[0.5mm]
	1+\frac{1-b}{\beta}
\end{array};-e^{\beta(\epsilon_{\text{max}}-\mu)}\right]\epsilon_{\text{max}}\nonumber\\[0.5mm]
	&+\frac{1+b\epsilon_{\text{max}}}{(b-1)^2}{}_3F_2\left[\begin{array}{c}
		1,\frac{1-b}{\beta},\frac{1-b}{\beta}\\[0.5mm]
		1+\frac{1-b}{\beta},1+\frac{1-b}{\beta}
	\end{array};-e^{\beta(\epsilon_{\text{max}}-\mu)}\right]
	-{}_3F_2\left[\begin{array}{c}
		1,\frac{1}{\beta},\frac{1}{\beta}\\[0.5mm]
		1+\frac{1}{\beta},1+\frac{1}{\beta}
	\end{array};-e^{\beta(\epsilon_{\text{max}}-\mu)}\right]\nonumber\\[0.5mm]
	&+\frac{2b}{(b-1)^2}{}_4F_3\left[\begin{array}{c}
		1,\frac{1-b}{\beta},\frac{1-b}{\beta},\frac{1-b}{\beta}\\[0.5mm]
		1+\frac{1-b}{\beta},1+\frac{1-b}{\beta},1+\frac{1-b}{\beta}
	\end{array};-e^{\beta(\epsilon_{\text{max}}-\mu)}\right]\Bigg\}
\end{alignat}
We can again take the limit $e^{\beta(\epsilon_{\text{max}}-\mu)}\gg 1$, dividing out $\langle E\rangle$, to obtain
\begin{equation}
	\frac{\langle U\rangle}{\langle E\rangle} \simeq
	\begin{cases}
		\mu-\frac{\pi}{\beta}\cot\left(\frac{\pi}{\beta}\right)&\quad\text{if}\quad\beta>1\\[0.5cm]
		\frac{1}{b}+\frac{1}{2}\epsilon_{\text{max}}+\frac{1}{2}\mu&\quad\text{if}\quad\beta=1\\[0.5cm]
		\epsilon_{\text{max}}-\frac{b+3 \beta -3}{(1-\beta) (b+\beta -1)}&\quad\text{if}\quad\beta<1.
	\end{cases}
\end{equation}
Finally, this leads us to the entropy:
\begin{equation}
	\frac{S}{\langle E\rangle}=\beta\left(\frac{\langle U\rangle}{\langle E\rangle} -\frac{\Xi}{\langle E\rangle}-\mu \right)\simeq\begin{cases}
		 \beta\left(\mu-\frac{\pi}{\beta}\cot\left(\frac{\pi}{\beta}\right)+1-\mu\right)&\quad\text{if}\quad\beta>1\\[0.5cm]
		 \frac{1}{b}+\frac{1}{2}\epsilon_{\text{max}}+\frac{1}{2}\mu+1-\mu&\quad\text{if}\quad\beta=1\\[0.5cm]
		 \beta\left(\epsilon_{\text{max}}-\frac{b+3\beta-3}{(1-\beta)(b+\beta-1)}+\frac{1}{\beta}-\mu\right)&\quad\text{if}\quad\beta<1
	\end{cases}\label{eq:Sgeneral}
\end{equation}
Now we plug in the remaining variables to obtain
\begin{alignat}{6}
	\frac{S}{\langle E\rangle }\simeq
	\begin{cases}
		\beta-\pi\cot\left(\frac{\pi}{\beta}\right)&\text{ if }\beta>1\\[0.5cm]
		1+\frac{1}{2}\ln N-\frac{1}{2}\ln \langle k\rangle+\ln\left(\frac{\gamma-1}{\gamma-2}\right)+\frac{1}{2-\gamma}+\frac{1}{2}\ln\ln N&\text{ if }\beta=1\\[0.5cm]
		1+\ln N-\ln \langle k\rangle -\ln(1-\beta)+\frac{\beta}{\beta -1} +2\ln\left(\frac{\gamma-1}{\gamma-2}\right) -\frac{2}{\gamma -2}&\text{ if }\beta<1
	\end{cases}
\end{alignat}
The final entropy is, as expected, equal to that of the an Erd\"os-Renyi graph with connection probability $\langle k\rangle/N$ when $\beta\rightarrow0$ and $\gamma\rightarrow\infty$\cite{Anand2014} and should give the entropy of the soft configuration model when $\beta\rightarrow 0$.  \\

\begin{figure}[h]
	\vspace{-5mm}
	\centering
	\includegraphics[width=0.8\linewidth]{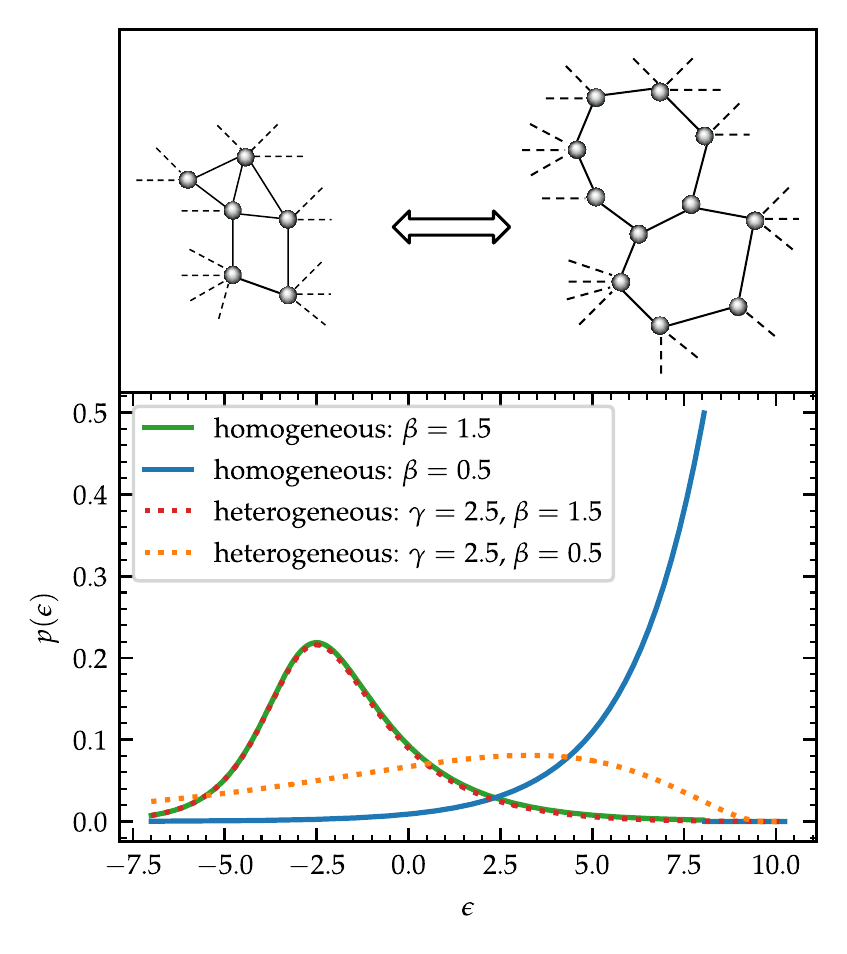}
	\vspace{-0.2cm}\caption{The probability $p(\epsilon)$ of finding a link with energy $\epsilon$ based on Eq. \eqref{eq:pe}. The full lines show the homogeneous case whereas the dotted lines represent the heterogeneous case with $\gamma=2.5$. For both degree distributions we plot the $p(\epsilon)$ for both $\beta=0.5$ (blue/orange) and $\beta=1.5$ (green/red). In all cases $N=10^5$ and $\langle k\rangle = 4$, i.e. this represents the situation for a sparse graph.}
	\label{fig:pe}
\end{figure}

Using the density of states and the Fermi-Dirac statistics we can also find the probability of a link having energy $\epsilon$. This is namely given by
\begin{equation}
	p(\epsilon)=\frac{1}{\langle E\rangle }\frac{g(\epsilon)}{1+e^{\beta(\epsilon-\mu)}}\label{eq:pe}
\end{equation}
We can plug in the approximated values of $\mu$ from before and plot the results for $\beta=1/2$ (blue line) and $\beta=3/2$ (orange line) in Fig. \ref{fig:pe}. Notice that this probability density changes dramatically at the ``critical'' point $\beta=1$. Indeed, when $\beta>1$ particles occupy low energy states and for $\beta<1$ they occupy mainly high energy states. However, since the number of states grows exponentially with the energy, the number of available microstates per particle grows extremely fast in the $\beta<1$ regime, inducing a sudden increase of the entropy, explaining the divergence of the entropy in the thermodynamic limit in this regime. 

\subsubsection{Toy model}
Above we have seen some interesting bahavior, most notably the non-extensivity of the entropy above the critical temperature. We want to now investigate where this feature comes from, by looking at a simplified version of our model. Suppose we have a system made of $N_{\text{part}}$ non-interacting ``particles'', each of which can attain states of energy $\epsilon \in (0,\epsilon_{\text{max}})$. Suppose also that the degeneracy of states of energy $\epsilon$ grows as
\[
g(\epsilon) = V e^{\beta_c \epsilon}
\]
with $\beta_c$ a fixed parameter and $V$ the volume of the system. The probability density to find one such particle in a state of energy $\epsilon$ is
\begin{equation}
	p(\epsilon)=\frac{\beta-\beta_c}{1-e^{-(\beta-\beta_c)\epsilon_{\text{max}}}}e^{-(\beta-\beta_c)\epsilon}.
	\label{p_epsilon}
\end{equation}
We notice that here we find the same sudden change of behavior at the critical point $\beta=\beta_c$ as we found in the $\mathbb{S}^1$ model. Using Maxwell-Boltzmann statistics for identical particles, the entropy per particle of this system is easily calculated as
\begin{equation}
	\frac{S}{N_{\text{\text{part}}}}=\frac{\beta}{\beta-\beta_c}-\frac{\beta \epsilon_{\text{max}}}{e^{(\beta-\beta_c)\epsilon_{\text{max}}}-1}-\ln{\left[ \frac{N_{\text{part}}}{V}\frac{\beta-\beta_c}{1-e^{-(\beta-\beta_c)\epsilon_{\text{max}}}}\right]}+1.
\end{equation}
The first two terms in this last equation are just the average energy per particle of the system. If the density of particles is kept fixed, so that $\lim_{N_{\text{part}} \rightarrow \infty}\frac{N_{\text{part}}}{V}=\text{cte}$, then entropy is an extensive quantity as it is proportional to the number of particles. However, there is a clear change of behavior as one goes from $\beta>\beta_c$ to $\beta<\beta_c$ due to the change of behavior of the probability density Eq.~\eqref{p_epsilon}. If besides $\beta \epsilon_{\text{max}} \gg 1$, then the entropy behaves as
\begin{equation}
	\frac{S}{N_{\text{\text{part}}}} \simeq
	\left\{
	\begin{array}{lr}
		\frac{\beta}{\beta-\beta_c}-\ln{\left[ \frac{N_{\text{part}}}{V}(\beta-\beta_c)\right]}+1 & \beta>\beta_c \\[0.5cm]
		\frac{1}{2} \beta_c \epsilon_{\text{max}} -\ln{\left[ \frac{N_{\text{part}}}{V\epsilon_{\text{max}}} \right]}+1& \beta=\beta_c \\[0.5cm]
		\beta_c \epsilon_{\text{max}}+\frac{\beta}{\beta-\beta_c}-\ln{\left[ \frac{N_{\text{part}}}{V}(\beta_c-\beta)\right]}+1 & \beta<\beta_c
	\end{array}
	\right.
	\label{eq:toyS}
\end{equation}
Thus, in the limit of $\epsilon_{\text{max}}\rightarrow \infty$ the entropy per particle diverges at $\beta \rightarrow \beta_c^+$ and scales as $\epsilon_{\text{max}}$ for $\beta \le \beta_c$, just as in our model.\\

 In the $\mathbb{S}^1$-model, the effective system size is given by the proportionality constant $V_{\text{eff}}=ae^{\epsilon_{\text{max}}}$ in Eq. \eqref{eq:dosgen} and the amount of particles is given by $\langle E\rangle=N\langle k\rangle/2$ as we are working with a sparse graph. In this case we indeed satisfy $\lim_{{\langle E\rangle} \rightarrow \infty}\frac{\langle E\rangle}{V_{\text{eff}}}=\text{cte}$ and the entropy is thus in principle extensive. However, as in the full model there is the extra constraint $\langle E\rangle\leq N(N-1)/2$ and $\epsilon_{\text{max}}=\ln(N/(2\kappa_0^2))$, we are obliged to also send $\epsilon_{\text{max}}$ to infinity when going to the thermodynamic limit, thus resulting in a non-extensive entropy for $\beta<\beta_c$. We thus show that the essential feature of the $\mathbb{S}^1$ model that leads to a non-extensive entropy is the exponential dependence on the energy of the density of states. \\

\subsection{Scaling Behaviour of Clustering with System Size}
In the following section we find the dominant finite size scaling of the clustering coefficient for $\beta\leq1$. As was explained in the main text, in this region in the thermodynamic limit clustering vanishes. We will therefore study what happens when $N\gg1$ but finite for any $\beta$ (we thus do not take any limit with respect to the temperature). As for $\beta\lesssim1$ higher order finite size correction become important, we study separately the case $\beta=1$. \\

We start by manipulating the angular integrals of Eq. \eqref{eq:clustering} as to simplify the task at hand later on. We then turn to the scaling when $\beta<1$ and conclude with an analysis of the scaling when $\beta=1$. In the case of $\beta<1$, in order to facilitate numerics later on, we choose to adopt the connection probability as defined by Eq. \eqref{eq:Fbetasmall}, where the degree sequence at different temperatures is the same. \\

The basis of these calculations is the fact that we are looking for the scaling behaviour of the $\overline{c}$ with respect to the system size $N$. This allows us to always ignore terms that we know are smaller than than the main term, which simplifies the integrals that we study substantially. Once we have a term, say $A$, we want to know the scaling behaviour of, we use the fact that if the functions $f(N)$ and $g(N)$ in equation
\begin{equation}
	f(N)<A<g(N)
\end{equation}
have the same dominant scaling, one can immediately conclude that $A$ also has that exact dominant scaling. Therefore, by finding upper and lower bounds to the integrals in question we can extract there scaling behavior with respect to $A$. It is important to keep in mind that, even when the integrals representing the bounds become very tedious, the strategy we employ remains the same throughout this section.

\subsubsection{Angular Manipulation}
We start by manipulating the angular integrals of Eq. \eqref{eq:clustering} to make it easier to work with, i.e. get rid of the absolute values in the expressions for $\Delta\theta$. The equation has the following form:
\begin{alignat}{6}
    \overline{c}&(\kappa)=
    &\frac{\iiiint\mathrm{d}\kappa'\mathrm{d}\kappa''\mathrm{d}\theta'\mathrm{d}\theta''\rho(\kappa')\rho(\kappa'')p(\kappa,\kappa',\pi-|\pi-|\theta'||)p(\kappa,\kappa'',\pi-|\pi-|\theta''||)p(\kappa',\kappa'',\pi-|\pi-|\theta'-\theta''||))}{\iint\mathrm{d}\kappa'\mathrm{d}\theta'\rho(\kappa')p(\kappa,\kappa',\pi-|\pi-|\theta'||)}.
\end{alignat}
Here, we have used $\theta=0$. Let us first investigate the trivial case of the denominator, where we only focus on the angular integral
\begin{alignat}{6}
    \int_0^{2\pi}\mathrm{d}\theta' p(\kappa,\kappa',\pi-|\pi-|\theta'||)&=\int_0^{\pi}\mathrm{d}\theta' p(\kappa,\kappa',\pi-|\pi-|\theta'||)+\int_{\pi}^{2\pi}\mathrm{d}\theta' p(\kappa,\kappa',\pi-|\pi-|\theta'||)\nonumber\\
    &=\int_0^{\pi}\mathrm{d}\theta' p(\kappa,\kappa',\theta')+\int_{\pi}^{2\pi}\mathrm{d}\theta' p(\kappa,\kappa',2\pi-\theta')=2\int_0^{\pi}\mathrm{d}\theta' p(\kappa,\kappa',\theta'),
\end{alignat}
where in the last step we have performed the transformation $t=2\pi-\theta'$ and $t\rightarrow\theta'$ on the second integral. 
The numerator can be rewritten in a similar way to obtain four terms
\begin{alignat}{6}
  &\int_0^{2\pi}\mathrm{d}\theta'\int_0^{2\pi}\mathrm{d}\theta''p(\kappa,\kappa',\pi-|\pi-|\theta'||)p(\kappa,\kappa'',\pi-|\pi-|\theta''||)p(\kappa',\kappa'',\pi-|\pi-|\theta'-\theta''||)\nonumber\\
  =&2\int_0^{\pi}\mathrm{d}\theta'\bigg(\int_0^{\theta'}\mathrm{d}\theta''p(\kappa,\kappa',\theta')p(\kappa,\kappa'',\theta'')p(\kappa',\kappa'',\theta'-\theta'')+\int_0^{\theta'}\mathrm{d}\theta''p(\kappa,\kappa',\theta'')p(\kappa,\kappa'',\theta')p(\kappa',\kappa'',\theta'-\theta'')\nonumber\\
  +&\int_0^{\pi-\theta'}\mathrm{d}\theta''p(\kappa,\kappa',\theta')p(\kappa,\kappa'',\theta'')p(\kappa',\kappa'',\theta'+\theta'')+\int_{\pi-\theta'}^{\pi}\mathrm{d}\theta''p(\kappa,\kappa',\theta')p(\kappa,\kappa'',\theta'')p(\kappa',\kappa'',2\pi-\theta'-\theta'')\bigg).
\end{alignat}
The first two terms are not exactly the same. However, as the full expression of the clustering coefficient also contains integrals over the hidden degrees, one can interchange $\kappa'\leftrightarrow\kappa''$. This thus shows that the first two terms contribute equally to the clustering coefficient. All in all, we will thus be working with the following three terms
\begin{alignat}{6}
    &4\int_0^{\pi}\mathrm{d}\theta'\int_0^{\theta'}\mathrm{d}\theta''p(\kappa,\kappa',\theta')p(\kappa,\kappa'',\theta'')p(\kappa',\kappa'',\theta'-\theta'')\nonumber\\
    +&2\int_{0}^{\pi}\mathrm{d}\theta'\int_0^{\pi-\theta'}\mathrm{d}\theta''p(\kappa,\kappa',\theta')p(\kappa,\kappa'',\theta'')p(\kappa',\kappa'',\theta'+\theta'')\nonumber\\
    +&2\int_{0}^{\pi}\mathrm{d}\theta'\int_{\pi-\theta'}^{\pi}\mathrm{d}\theta''p(\kappa,\kappa',\theta')p(\kappa,\kappa'',\theta'')p(\kappa',\kappa'',2\pi-\theta'-\theta'').\label{eq:anglesfinal}
\end{alignat}
Now, before we get started on finding the scaling with respect to the system size of each term individually, it might be that we can avoid doing so by some simple arguments. Indeed, we will show that the first term will always dominate the others in the large $N$ limit, and so we only have to find its scaling.
Let us start with the second term
\begin{alignat}{6}
    &2\iint\mathrm{d}\kappa'\mathrm{d}\kappa''\rho(\kappa')\rho(\kappa'')\int_{0}^{\pi}\mathrm{d}\theta'\int_0^{\pi-\theta'}\mathrm{d}\theta''p(\kappa,\kappa',\theta')p(\kappa,\kappa'',\theta'')p(\kappa',\kappa'',\theta'+\theta'')\nonumber\\
    \leq&2\iint\mathrm{d}\kappa'\mathrm{d}\kappa''\rho(\kappa')\rho(\kappa'')\int_{0}^{\pi}\mathrm{d}\theta'\int_0^{\pi}\mathrm{d}\theta''p(\kappa,\kappa',\theta')p(\kappa,\kappa'',\theta'')p(\kappa',\kappa'',\theta'+\theta'').
\end{alignat}
The above statement is true as the integrand is strictly positive and so extending the integration domain will only make the integral larger. Now, we can split the $\theta''$ integral and perform $\theta'\leftrightarrow\theta''$ and $\kappa'\leftrightarrow\kappa''$ on the second term to obtain
\begin{alignat}{6}
  2\iint\mathrm{d}\kappa'\mathrm{d}\kappa''\rho(\kappa')\rho(\kappa'')&\int_{0}^{\pi}\mathrm{d}\theta'\int_0^{\pi}\mathrm{d}\theta'' p(\kappa,\kappa',\theta')p(\kappa,\kappa'',\theta'')p(\kappa',\kappa'',\theta'+\theta'')\nonumber\\
  &=4\iint\mathrm{d}\kappa'\mathrm{d}\kappa''\rho(\kappa')\rho(\kappa'')\int_{0}^{\pi}\mathrm{d}\theta'\int_0^{\theta'}\mathrm{d}\theta''p(\kappa,\kappa',\theta')p(\kappa,\kappa'',\theta'')p(\kappa',\kappa'',\theta'+\theta'')\nonumber\\
  &\leq4\iint\mathrm{d}\kappa'\mathrm{d}\kappa''\rho(\kappa')\rho(\kappa'')\int_{0}^{\pi}\mathrm{d}\theta'\int_0^{\theta'}\mathrm{d}\theta''p(\kappa,\kappa',\theta')p(\kappa,\kappa'',\theta'')p(\kappa',\kappa'',\theta'-\theta'').
\end{alignat}
In the final step we use the functional form of $p$ with respect to the angular coordinate is
\begin{equation}
    p(s)=\frac{1}{1+s^\beta}.\label{eq:funcformF}
\end{equation}
As $s^\beta$ is monotonously increasing, and $1/(1+s)$ is monotonously decreasing, $p(s)$ is monotonously decreasing. Thus, it is largest when $s$ is smallest. Obviously, $\theta'+\theta''>\theta'-\theta''$ for all $(\theta',\theta'')\in[0,\pi]\times[0,\theta']$. We have thus proven that the first term in Eq. \eqref{eq:anglesfinal} dominates the second term. We can follow similar steps for the third term. We we will now only clarify steps if they are new. 
\begin{alignat}{6}
    &2\iint\limits_{\kappa',\kappa''}\mathrm{d}\kappa'\mathrm{d}\kappa''\rho(\kappa')\rho(\kappa'')\int\limits_{0}^{\pi}\mathrm{d}\theta'\int\limits_{\pi-\theta'}^\pi\mathrm{d}\theta''p(\kappa,\kappa',\theta')p(\kappa,\kappa'',\theta'')p(\kappa',\kappa'',2\pi-\theta'-\theta'')\nonumber\\
    \leq&4\iint\limits_{\kappa',\kappa''}\mathrm{d}\kappa'\mathrm{d}\kappa''\rho(\kappa')\rho(\kappa'')\int\limits_{0}^{\pi}\mathrm{d}\theta'\int\limits_0^{\theta'}\mathrm{d}\theta''p(\kappa,\kappa',\theta')p(\kappa,\kappa'',\theta'')p(\kappa',\kappa'',2\pi-\theta'-\theta'').
\end{alignat}
Now, one knows that $2\pi-\theta'-\theta''\geq\theta'-\theta''$ $\forall_{(\theta',\theta'')\in[0,\pi]\times[0,\theta']}$. For the same reasons as before, this then implies
\begin{alignat}{6}
   &4\iint\limits_{\kappa',\kappa''}\mathrm{d}\kappa'\mathrm{d}\kappa''\rho(\kappa')\rho(\kappa'')\int\limits_{0}^{\pi}\mathrm{d}\theta'\int\limits_0^{\theta'}\mathrm{d}\theta''p(\kappa,\kappa',\theta')p(\kappa,\kappa'',\theta'')p(\kappa',\kappa'',2\pi-\theta'-\theta'')\nonumber\\
   \leq &4\iint\limits_{\kappa',\kappa''}\mathrm{d}\kappa'\mathrm{d}\kappa''\rho(\kappa')\rho(\kappa'')\int\limits_{0}^{\pi}\mathrm{d}\theta'\int\limits_0^{\theta'}\mathrm{d}\theta''p(\kappa,\kappa',\theta')p(\kappa,\kappa'',\theta'')p(\kappa',\kappa'',\theta'-\theta''),
\end{alignat}
so this term is also dominated by the first term in Eq. \eqref{eq:anglesfinal}. 

\subsubsection{Case $0<\beta<1$}
The first step is to perform the transformation $x=\frac{\kappa'}{\kappa_s}$ and $y=\frac{\kappa''}{\kappa_s}$, where we define $\kappa_s^2\equiv N^\beta/((2\pi)^\beta\hat{\mu})$. Note that we use assume the functional form of $\hat{\mu}$ defined in Eq. \eqref{eq:muhightemp}, such that $\kappa_s\sim \sqrt N$. This leads to
\begin{alignat}{6}
    \overline{c}(\kappa)\sim \,&2\,\frac{\int\limits^{\kappa_c/\kappa_s}_{\kappa_0/\kappa_s}\mathrm{d} x\int\limits^{\kappa_c/\kappa_s}_{\kappa_0/\kappa_s}\mathrm{d}y\int\limits^{\pi}_0\mathrm{d}\theta'\int\limits^{\theta'}_0 \mathrm{d}\theta''  (xy)^{-\gamma}p(\kappa,\kappa_sx,\theta')p(\kappa,\kappa_sy,\theta'')p(\kappa_sx,\kappa_sy,\theta'-\theta'')}{\left(\int\limits^{\kappa_c/\kappa_s}_{\kappa_0/\kappa_s}\mathrm{d}x\int\limits^{\pi}_0\mathrm{d}\theta' x^{-\gamma}p(\kappa,\kappa_sx,\theta')\right)^2}\label{eq:fullC},
\end{alignat}

We investigate the numerator and denominator separately and define
\begin{alignat}{6}
    A_-&=\int\limits^{\kappa_c/\kappa_s}_{\kappa_0/\kappa_s}\mathrm{d}x\int\limits^{\kappa_c/\kappa_s}_{\kappa_0/\kappa_s}\mathrm{d}y\int\limits^{\pi}_0\mathrm{d}\theta'\int\limits^{\theta'}_0 \mathrm{d}\theta''  (xy)^{-\gamma}p(\kappa,\kappa_sx,\theta')p(\kappa,\kappa_sy,\theta'')p(\kappa_sx,\kappa_sy,\theta'-\theta'').\\
    B&=\int\limits^{\kappa_c/\kappa_s}_{\kappa_0/\kappa_s}\mathrm{d}x\int\limits^{\pi}_0\mathrm{d}\theta' x^{-\gamma}p(\kappa,\kappa_sx,\theta').
\end{alignat}
It is also useful to define
\begin{equation}
    A_+=\int\limits^{\kappa_c/\kappa_s}_{\kappa_0/\kappa_s}\mathrm{d}x\int\limits^{\kappa_c/\kappa_s}_{\kappa_0/\kappa_s}\mathrm{d}y\int\limits^{\pi}_0\mathrm{d}\theta'\int\limits^{\theta'}_0 \mathrm{d}\theta''  (xy)^{-\gamma}p(\kappa,\kappa_sx,\theta')p(\kappa,\kappa_sy,\theta'')p(\kappa_sx,\kappa_sy,\theta'+\theta'').\label{eq:defA+}
\end{equation}
Our investigation will focus on finding upper and lower bounds for these integrals. Note that from here on out we will drop the domains of the $x$ and $y$ integrals and assume them to be $[\kappa_0/\kappa_s,\kappa_c/\kappa_s]$ unless otherwise indicated. Using the fact that 
\begin{equation}
    \frac{1}{1+\dfrac{(\theta'+\theta'')^\beta}{xy}}<\frac{1}{1+\dfrac{(\theta'-\theta'')^\beta}{xy}}, \quad \forall_{ \theta',\theta'',x,y},
\end{equation}
we can conclude that $A_+<A_-$. As numerical investigation leads us to expect that both have the same scaling, this implies that we do not need to worry about an upper bound for $A+$ nor the lower bound for $A_-$. If the functions $f(N)$ and $g(N)$ in equation
\begin{equation}
    f(N)<A_+<A_-<g(N)
\end{equation}
have the same dominant scaling, one can immediately conclude that $A_-$ also has that exact dominant scaling. One might ask why we introduce $A_+$ in the first place, when in the end we are only interested in the scaling of $A_-$. The answer to this is that $A_+$ in general has nicer properties due to the lack of $(\theta'-\theta'')$, as it is thus easier to find a lower bound for it than for $A_-$.\\

We start with the simplest integral, the $B$-term, which can be solved exactly. To this end we first need to rewrite it a bit. By performing two substitutions 
\begin{equation}
    x'=\frac{\kappa_s}{\kappa_c}x \hspace{5mm} x'\rightarrow x, \hspace{5mm} t=\frac{\theta'}{\pi} \hspace{5mm} t\rightarrow \theta',
\end{equation}
one obtains
\begin{equation}
    B=\pi\left(\frac{\kappa_c}{\kappa_s}\right)^{1-\gamma}\int\limits^1_0\mathrm{d}\theta'\int\limits^1_{\kappa_0/\kappa_c}\mathrm{d}x\frac{x^{-\gamma}}{1+\dfrac{(\pi\theta')^\beta\kappa_s^2}{\kappa_c\kappa x}}.
\end{equation}
This then gives the following expression
\begin{alignat}{8}
    B=\frac{\pi}{(\beta(\gamma-1)-1)(\gamma-1)}\Bigg\{&(\gamma-1)\beta&&\left(\frac{\kappa_0}{\kappa_s}\right)^{1-\gamma}&&_2F_1\left[\begin{array}{c}1,1/\beta\\1+1/\beta\end{array};-\frac{\pi^\beta\kappa_s^2}{\kappa\kappa_0}\right]\nonumber\\
    -&(\gamma-1)\beta&&\left(\frac{\kappa_c}{\kappa_s}\right)^{1-\gamma}&&_2F_1\left[\begin{array}{c}1,1/\beta\\1+1/\beta\end{array};-\frac{\pi^\beta\kappa_s^2}{\kappa\kappa_0}\right]\nonumber\\
    -&\quad\,\,\kappa_0&&\left(\frac{\kappa_0}{\kappa_s}\right)^{1-\gamma}&&_2F_1\left[\begin{array}{c}1,\gamma-1\\\gamma\end{array};-\frac{\pi^\beta\kappa_s^2}{\kappa\kappa_0}\right]\nonumber\\
    +&\quad\,\,\kappa_0&&\left(\frac{\kappa_c}{\kappa_s}\right)^{1-\gamma}&&_2F_1\left[\begin{array}{c}1,\gamma-1\\\gamma\end{array};-\frac{\pi^\beta\kappa_s^2}{\kappa\kappa_0}\right]\Bigg\}.\label{eq:Bfull}
\end{alignat}
This expression can then be expanded w.r.t. $N$, using that $\kappa_s\sim \sqrt{N}$. To lowest order, one finds that $B$ then scales as
\begin{equation}
    B\sim N ^{\frac{\gamma-3}{2}}
\end{equation}

Next we turn to the $A_+$ term. Here we use the following fact to bound this integral. If $F=\int_\mathcal{V} f(\vec{x})$, where $\mathcal{V}$ is the volume over which to integrate the function, and $f(\vec{x})\geq f_0$ for all $\vec{x}$, where $f_0$ some constant, then $F\geq f_0\mathcal{V}$. From the form of the standard connection probability given in Eq. \eqref{eq:funcformF}, we see that $A_+$ is smallest when the argument is largest, which is the case when $\theta',\theta''$ are largest, so when they are both $\pi$. Thus we can bound the angular integrals by replacing the integrand with its minimum, the same function where both angular coordinates are $\pi$. The integrand is then a constant so the bound is given by the value of that constant times the area of the integral. Plugging this in we obtain
\begin{alignat}{6}
    A_+&\geq \frac{1}{2}\pi^2\int\limits\mathrm{d} x\mathrm{d} y (xy)^{-\gamma}\frac{1}{1+\dfrac{\pi^\beta\kappa_s}{\kappa x}}\frac{1}{1+\dfrac{\pi^\beta\kappa_s}{\kappa y}}\frac{1}{1+\dfrac{(2\pi)^\beta}{xy}}\nonumber\\
    &= \frac{1}{2}\pi^{2-3\beta}\left(\frac{\kappa}{\kappa_s}\right)^2\int\limits\mathrm{d} x\mathrm{d} y (xy)^{2-\gamma}\frac{1}{1+\dfrac{\kappa x}{\pi^\beta\kappa_s}}\frac{1}{1+\dfrac{\kappa y}{\pi^\beta\kappa_s}}\frac{1}{1+\dfrac{xy}{(2\pi)^\beta}}.
\end{alignat}
Now, this is exactly the same integral (with the exception of the $\pi$'s, but they will obviously not change scaling) as the one evaluated in Ref. \cite{colomer-de-simon2012clustering}. As was found in the reference (Eq. (6)), the scaling depends on how we set $\kappa_c$ relative to $\kappa_s$. We distinguish two regimes. First, there is the regime where $\kappa_0\ll\kappa_s\ll\kappa_c$. In this case, the scaling is
\begin{equation}
    A_+\geq c_{+,1}\kappa_s^{-2}\ln(\kappa_c/\kappa_s).\label{eq:A+1term1}
\end{equation}
Then, there is the region where $\kappa_0\leq\kappa_c\leq\kappa_s$ ($\kappa_0\ll\kappa_s$ must be required to hold) where one obtains
\begin{equation}
    A_+\geq c_{+,2}\kappa_s^{2\gamma-8}\kappa_c^{6-2\gamma}.
\end{equation}
This, however, does not give the full scaling behaviour, as numerical results show us that for large $\beta$ the scaling with respect to $N$ is different. To find where this different scaling comes from we take a step back and look at the full integral $A_+$ as given in Eq. \eqref{eq:defA+}. One might be tempted to, as in Ref. \cite{colomer-de-simon2012clustering}, expand the first two connection probabilities to first order. However, the presence of the angular coordinate makes this impossible. The argument of these connection probabilities has the form $s=\frac{\theta^\beta \kappa_s^2}{\kappa \kappa'}$. It becomes clear that for small enough $\theta$, $s$ is no longer large and the approximation thus breaks down. We thus expect different scaling behaviour to arise as a result of small angular coordinates. To investigate this further, we split the \textit{angular} integration domain $[0,\pi]\times[0,t]$ in a convenient way and investigate the domain  $\mathcal{D}_1=[0,(xy)^{1/\beta}]\times[0,t]$. Note that we do not have to look at the other half of the original domain as we are only interested in the lower bound and our integrand is positive for all angles, which means that the integral over the full domain must be larger or equal to the integral over $D_{1}$. The domain $\mathcal{D}_1$ can only be defined in the case that $\kappa_c\leq\kappa_s$, as only then the angular coordinates remain smaller than the maximal possible value of $\pi$ for all $x$ and $y$. For the case that $\kappa_c\gg\kappa_s$ we define the more restrictive domain $\mathcal{D}_2=[0,(\kappa_0/\kappa_s)^{2/\beta}]\times[0,t]$. Starting with the case $\kappa_c\leq\kappa_s$, bounding the integral as before (by replacing the integrand by its minimum), one finds
\begin{alignat}{6}
    A_+&\geq\frac{1}{1+2^\beta} \int\limits\mathrm{d} x\mathrm{d} y (xy)^{2/\beta-\gamma}\frac{1}{1+\dfrac{\kappa_s y}{\kappa }}\frac{1}{1+\dfrac{\kappa_s x}{\kappa }}\nonumber\\
    &= \frac{(\kappa_s/\kappa)^{-4/\beta+2\gamma
   -2} }{1+2^\beta}\left( B_{\frac{\kappa }{\kappa_0+\kappa}}\left[\gamma -\frac{2}{\beta},1-\gamma+\frac{2}{\beta}\right]-B_{\frac{\kappa }{\kappa_c+\kappa}}\left[\gamma -\frac{2}{\beta},1-\gamma+\frac{2}{\beta}\right]\right)^2\nonumber\\[2mm]
   &\simeq c_{+,s,1}\kappa_s^{-4/\beta+2\gamma-2}+c_{+,s,2}\kappa_s^{-4/\beta+2\gamma-2}\kappa_c^{4/\beta-2\gamma},
\end{alignat}
For the case $\kappa_c\gg\kappa_s$ one obtains
\begin{alignat}{6}
    A_+&\geq\left(\dfrac{\kappa_0}{\kappa_s}\right)^{4/\beta} \int\limits\mathrm{d} x\mathrm{d} y (xy)^{-\gamma}\frac{1}{1+\dfrac{\kappa_s}{\kappa x}\dfrac{\kappa_0^2}{\kappa_s^2}}\frac{1}{1+\dfrac{\kappa_s}{\kappa y}\dfrac{\kappa_0^2}{\kappa_s^2}}\frac{1}{1+\dfrac{2^\beta}{xy}\dfrac{\kappa_0^2}{\kappa_s^2}}\nonumber\\
    &\simeq \left(\dfrac{\kappa_0}{\kappa_s}\right)^{4/\beta} \int\limits\mathrm{d} x\mathrm{d} y (xy)^{-\gamma}\simeq c_{+,s,3}\kappa_s^{-4/\beta+2\gamma-2},\label{eq:A+1term2}
\end{alignat}
where in the first step it was noted that irrespective of the value of $x$ and $y$, the argument of the connection probabilities is small.

We now have five different scaling behaviours. Which terms dominate will depend on the value of $\beta$ as well on $\kappa_c$. To quantify how the scaling varies with $\kappa_c$ we introduce the exponent $\alpha$ such that $\kappa_c\sim N^{\alpha/2}$. As $\kappa_s\sim N^{1/2}$, the different regimes of $\kappa_c$ described above correspond to $\alpha\in[0,1]$ for $\kappa_c\leq\kappa_s$ and $\alpha\in(1,\frac{2}{\gamma-1}]$ for $\kappa_c\gg\kappa_s$. Using these definitions and adding up the different scaling we found above, we conclude that
\begin{equation}
    A_+\geq\begin{cases}
     C_{+,1} N^{-2/\beta+\gamma-1}+C_{+,2} N^{-1}\ln N\quad &\text{if}\quad \kappa_c\gg\kappa_s\\
     N^{-1}\bigg(C_{+,3}N^{\gamma-2/\beta}+ C_{+,4}N^{(1-\alpha)(\gamma-2/\beta)}+C_{+,5} N^{(1-\alpha)(\gamma-3)}\bigg)\quad &\text{if}\quad \kappa_c\leq\kappa_s
    \end{cases}\label{eq:A+lowerbound} ,
\end{equation}
where $C_{+,i}$ are constants. Note that, for example, the scaling of Eqs. \eqref{eq:A+1term1} and \eqref{eq:A+1term2} can indeed be combined to the first of these two inequalities as both now hold for all $\beta$. When $\beta>2/\gamma$ the $C_{+,2}$-term vanished with respect to the $C_{+,1}$-term and we are left with inequality \eqref{eq:A+1term2} and when $\beta<2/\gamma$ the other term dominates and we are left with inequality \eqref{eq:A+1term1}. \\

\begin{figure}[h]
	\centering
	\begin{tikzpicture}
		\def\AA{2.5}
		\fill[gray!50] (-0.8*\AA,-1*\AA) -- (-0.8*\AA,-0.8*\AA) -- (-0.4*\AA,-0.4*\AA) -- (1*\AA,-0.4*\AA) -- (1*\AA,-1*\AA) -- cycle;
		\fill[black] (-1*\AA,-1*\AA) -- (-0.8*\AA,-1*\AA) -- (-0.8*\AA,-0.8*\AA) --cycle;
		\fill[pattern=horizontal lines] (-0.4*\AA,-0.4*\AA) -- (1*\AA,-0.4*\AA) -- (\AA,\AA) -- cycle;
		\draw[black, very thick] (-1*\AA,-1*\AA) rectangle (1*\AA,1*\AA);
		\draw[black,very thick] (-1*\AA,-1*\AA) -- (1*\AA,1*\AA);
		\draw[black,very thick] (-0.8*\AA,-1*\AA) -- (-0.8*\AA,-0.8*\AA);
		\draw[dashed,very thick] (-0.4*\AA,-0.4*\AA) -- (1*\AA,-0.4*\AA);
		\draw[dashed,very thick] (-0.4*\AA,-1*\AA) -- (-0.4*\AA,-0.4*\AA);
		
		\draw[->] (-1.2*\AA,0.8*\AA) -- (-1.2*\AA,1*\AA);
		\node at (-1.2*\AA,0.7*\AA) { $\theta''$}; 
		\draw[->] (0.8*\AA,-1.2*\AA) -- (1*\AA,-1.2*\AA);
		\node at (0.7*\AA,-1.2*\AA) { $\theta'$};
		\node[above] at (-0.8*\AA,-1.2*\AA) { $b$};
		\node[above] at (-0.4*\AA,-1.2*\AA) { $a$};
		\node[left] at (1.2*\AA,-0.4*\AA) { $a$};
		\node at (0.4*\AA, 0*\AA) { $\mathcal{D}_X$};
		\node at (0.4*\AA, -0.8*\AA) { $\mathcal{D}_Y$};
		\node at (-0.6*\AA, -0.8*\AA) { $\mathcal{D}_Z$};
	\end{tikzpicture}%
	
	\caption{Integration regions. In the grey region ($\mathcal{D}_Y+\mathcal{D}_Z$) the argument of the Lerch zeta function is bigger than one, in the hatched region ($\mathcal{D}_X$) it is not and the black region is $\mathcal{D}_s$.}
	\label{fig:integrationregions}
\end{figure}
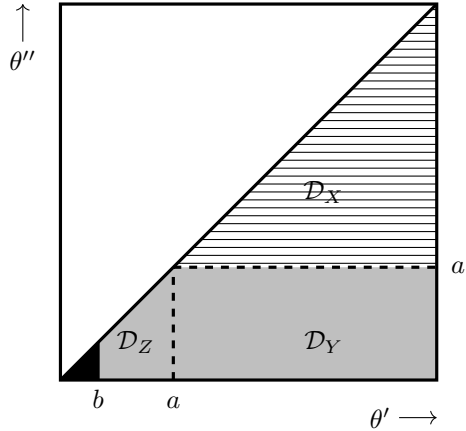

Now obviously this is just a lower bound. To show that the clustering indeed scales like this we must also find an upper bound, which we do by turning to the $A_-$ term. We divide the integration domain in two: $\mathcal{D}_s=[0,(\kappa_0/\kappa_s)^{2/\beta}]\times[0,\theta']$ and $\mathcal{D}_l=[(\kappa_0/\kappa_s)^{2/\beta},\pi]\times[0,\theta']$. We first turn to region $\mathcal{D}_l$. 
\begin{alignat}{6}
    A_{-,l}&= \iint_{\mathcal{D}_l}\mathrm{d}\theta'\mathrm{d}\theta''\iint\mathrm{d} x\mathrm{d} y (xy)^{-\gamma}&&\frac{1}{1+\dfrac{\theta'^\beta\kappa_s}{\kappa x}}\frac{1}{1+\dfrac{\theta''^\beta\kappa_s}{\kappa y}}\frac{1}{1+\dfrac{(\theta'-\theta'')^\beta}{xy}}\nonumber\\
    &\leq\left(\frac{\kappa}{\kappa_s}\right)^2\iint_{\mathcal{D}_l}\mathrm{d}\theta'\mathrm{d}\theta''\iint\mathrm{d} x\mathrm{d} y&& (xy)^{2-\gamma}(\theta'\theta''(\theta'-\theta''))^{-\beta}\frac{1}{1+\frac{xy}{(\theta'-\theta'')^{\beta}}}\nonumber\\
    &=\left(\frac{\kappa}{\kappa_s}\right)^2\iint_{\mathcal{D}_l}\frac{\mathrm{d}\theta'\mathrm{d}\theta''}{(\theta'\theta''(\theta'-\theta''))^{\beta}}&&\left(\frac{\kappa_c}{\kappa_s}\right)^{2(3-\gamma)}\Phi\left[-(\theta'-\theta'')^{-\beta}\left(\frac{\kappa_c}{\kappa_s}\right)^2,2,3-\gamma\right]\nonumber\\
    &+\left(\frac{\kappa}{\kappa_s}\right)^2\iint_{\mathcal{D}_l}\frac{\mathrm{d}\theta'\mathrm{d}\theta''}{(\theta'\theta''(\theta'-\theta''))^{\beta}}&&\left(\frac{\kappa_0}{\kappa_s}\right)^{2(3-\gamma)}\Phi\left[-(\theta'-\theta'')^{-\beta}\left(\frac{\kappa_0}{\kappa_s}\right)^2,2,3-\gamma\right]\nonumber\\
    &-2\left(\frac{\kappa}{\kappa_s}\right)^2\iint_{\mathcal{D}_l}\frac{\mathrm{d}\theta'\mathrm{d}\theta''}{(\theta'\theta''(\theta'-\theta''))^{\beta}}&&\left(\frac{\kappa_0\kappa_c}{\kappa_s^2}\right)^{3-\gamma}\Phi\left[-(\theta'-\theta'')^{-\beta}\frac{\kappa_0\kappa_c}{\kappa_s^2},2,3-\gamma\right].\label{eq:A-l3}
\end{alignat}
One sees that these three terms are similar, and so we treat the general integral
\begin{alignat}{6}
    I_\zeta&=\iint_{\mathcal{D}_l}\mathrm{d}\theta'\mathrm{d}\theta''\frac{\Phi\left[-(\theta'-\theta'')^{-\beta}\zeta,2,3-\gamma\right]}{(\theta'\theta''(\theta'-\theta''))^{\beta}}\zeta^{3-\gamma}=\iint_{\mathcal{D}_l}\mathrm{d}\theta'\mathrm{d}\theta''\frac{\Phi\left[-\theta''^{-\beta}\zeta,2,3-\gamma\right]}{(\theta'\theta''(\theta'-\theta''))^{\beta}}\zeta^{3-\gamma},
\end{alignat}
where the transformation $\theta'''=\theta'-\theta''$, $\theta'''\rightarrow\theta''$ was performed. Now, the argument of the Lerch zeta function can in principle be smaller and larger than one. If it is smaller, it can be shown that $\Phi[-(\theta'-\theta'')^{-\beta} \zeta,2,3-\gamma]<2^{\gamma-3}$. If it is bigger than one can use the identity described in Ref. \cite{colomer-de-simon2012clustering}
\begin{equation}
    \Phi[-z^2,2,3-\gamma]=z^{-2(3-\gamma)}\bigg(2\psi(\gamma)\ln z+\vartheta(\gamma)\bigg)+\frac{1}{z^2}\Phi\left[\frac{1}{z^2},2,\gamma-2\right]\label{eq:LerchPhiTransform},
\end{equation} 
where
\begin{equation}
    \psi(\gamma)=\Phi[-1,1,3-\gamma]+\Phi[-1,1,\gamma-2]\hspace{5mm}\text{and}\hspace{5mm}\vartheta(\gamma)=-\pi^2\cot(\pi\gamma)\csc(\pi\gamma).
\end{equation}
The argument of the Lerch zeta function is exactly one, which is the inflection point between the behaviours, when
\begin{equation}
    a=\zeta^{1/\beta}\label{eq:transitionpoint}
\end{equation}
We must thus split the integration domain $\mathcal{D}_l$ in three regions (where $b=(\kappa_0/\kappa_s)^{2/\beta}$): $\mathcal{D}_X=[a,\pi]\times[a,\theta']$, $\mathcal{D}_Y=[a,\pi]\times[0,a]$ and $\mathcal{D}_Z=[b,a]\times[0,\theta']$ as depicted in Fig. \ref{fig:integrationregions}.
Now, the grey region is the one where the Lerch zeta function argument is bigger than one, in the hatched region we can bound the Lerch zeta function away and the black region is $\mathcal{D}_s$ and we thus do not care about it for the moment. Before going any further, let us note that Fig. \ref{fig:integrationregions} looks slightly different for different $\kappa_c$ and $\zeta$. If $\kappa_c\gg\kappa_s$ and $\zeta=(\kappa_c/\kappa_s)^{2}$, then $\zeta\gg 1$ and thus so is $a$. However, $a$ as an integration limit must be smaller than $\pi$ and thus in this case the $\mathcal{D}_X$ and $\mathcal{D}_Y$ regions disappear. When $\kappa_c\leq\kappa_s$ this is not the case as for all $\zeta$, $a<\pi$. Finally, irrespective of the value of $\kappa_c$, for $\zeta=(\kappa_0/\kappa_s)^2$, $a=b$ and thus region $\mathcal{D}_Z$ vanishes. Implementing the transformation given by Eq. \eqref{eq:LerchPhiTransform} in the grey region one obtains
\begin{equation}
    I_\zeta\leq\iint\mathrm{d}\theta'\mathrm{d}\theta''(\theta'(\theta'-\theta''))^{-\beta}\Bigg\{\theta''^{\beta(2-\gamma)}\Big[\psi(\gamma)\ln\left(\frac{\zeta}{\theta''^{\beta}}\right)+\vartheta(\gamma)\Big]
    +\zeta^{2-\gamma}(\gamma-2)^{-2}\Bigg\}.
\end{equation}
As this leads to three different angular integrals, in the end we have seven different integrals to solve.
\begin{alignat}{6}
&\iint_{\mathcal{D}_X}\mathrm{d}\theta'\mathrm{d}\theta''\left(\frac{1}{\theta'\theta''(\theta'-\theta'')}\right)^{\beta}&&=\frac{a^{2-3\beta}}{3\beta-2}\bigg\{B_{1}[2\beta-1,1-\beta]-B_{1}[1-\beta,1-\beta]\nonumber\\[2mm]
&&&+B_{\frac{a}{\pi}}[2\beta-1,1-\beta]+(a/\pi)^{3\beta-2}B_{\frac{a}{\pi}}[1-\beta,1-\beta]\bigg\}\nonumber\\[2mm]
&&&+\frac{4^{\beta-1/2}\pi^{5/2-3\beta}\Gamma[1-\beta]}{\Gamma[3/2-\beta](3\beta-2)}\left((a/\pi)^{2-3\beta}-1\right)\\[2mm]
&&&=\boxed{c_{X_{11}} a^{2-3\beta}+c_{X_{12}}}\\[5mm]
%
&\iint_{\mathcal{D}_Y}\mathrm{d}\theta'\mathrm{d}\theta''\left(\frac{1}{\theta'(\theta'-\theta'')}\right)^{\beta}&&=\frac{a^{2-2\beta}}{2(\beta-1)^2}\bigg\{2(\beta-1)B_{\frac{a}{\pi}}[2\beta-1,1-\beta]\nonumber\\[2mm]
&&&-\pi^{-1/2}(\beta-1)\Gamma[1-\beta]\Gamma[\beta-1/2]-1\nonumber\\[2mm]
&&&+(1-{ }_2F_1\left[\begin{array}{c}2(\beta-1),\beta\\2\beta-1\end{array};a/\beta\right])(a/\pi)^{2\beta-2}\bigg\}\\[2mm]
&&&\simeq \boxed{c_{Y_{11}}a^{2-2\beta}+c_{Y_{12}}}\\[5mm]
%
&\iint_{\mathcal{D}_Y}\mathrm{d}\theta'\mathrm{d}\theta''\left(\frac{\theta''^{2-\gamma}}{\theta'(\theta'-\theta'')}\right)^{\beta}&&=\frac{a^{2-\gamma\beta}}{\gamma\beta-2}\bigg\{B_1[1+2\beta-\gamma\beta,1-\beta]\nonumber\\[2mm]
&&&-B_1[2\beta-1,1-\beta]+B_{\frac{a}{\pi}}[2\beta-1,1-\beta]\nonumber\\[2mm]
&&&-(a/\pi)^{\gamma\beta-2}B_{\frac{a}{\pi}}[1+2\beta-\gamma\beta,1-\beta]\bigg\}\\[2mm]
&&&\simeq\boxed{c_{Y_{21}}a^{2-\gamma\beta}+c_{Y_{22}}a^{1+2\beta-\gamma\beta}}\\[5mm]
&\iint_{\mathcal{D}_Y}\mathrm{d}\theta'\mathrm{d}\theta''\left(\frac{\theta''^{2-\gamma}}{\theta'(\theta'-\theta'')}\right)^{\beta}\ln\left(\frac{\zeta}{\theta''^{\beta}}\right)&&=\frac{\beta  a^{2-\beta  \gamma }\pi^{1-2\beta}}{(\beta  (\gamma -2)-1) (\beta  \gamma -2)^2}\nonumber\\[2mm]
&&&\times\bigg\{\frac{4^{\beta -1}}{ \pi ^{\frac{3}{2}-2 \beta }} (\beta  (\gamma -2)-1) \Gamma[1-\beta] \Gamma \left[\beta -\frac{1}{2}\right]\nonumber\\[2mm]
&&&+\frac{\pi ^{2 \beta -1} \Gamma [1-\beta ] \Gamma [-\gamma  \beta +2 \beta +2]}{\Gamma [-\gamma  \beta +\beta +2]}\nonumber\\[2mm]
&&&\times\left(\frac{1+(\gamma\beta-2)(H_{\beta(2-\gamma)}-H_{1+\beta-\gamma\beta})}{\beta(\gamma-2)-1}\right)\nonumber\\[2mm]
&&&-a^{2\beta-1}\left(\frac{(\beta  (\gamma -2)-1) }{2 \beta -1}\, _2F_1\left[\begin{array}{c}\beta ,2 \beta -1\\2 \beta \end{array};\frac{a}{\pi }\right]\right.\nonumber\\[2mm]
&&&\frac{(\beta  \gamma -2) \, }{\beta 
   (\gamma -2)-1}{}_3F_2\left[\begin{array}{c}
   \beta ,-\gamma  \beta +2 \beta +1,-\gamma  \beta +2 \\\beta +1-\gamma  \beta +2 \beta +2,-\gamma  \beta +2 \beta +2\end{array};\frac{a}{\pi }\right]\nonumber\\[2mm]
&&&+\, _2F_1\left[\begin{array}{c}\beta ,\beta  (-\gamma )+2 \beta +1\\\beta  (-\gamma )+2 \beta +2\end{array};\frac{a}{\pi }\right]\Bigg)\bigg\}\\[2mm]
&&&\simeq\boxed{c_{Y_{31}}a^{2-\gamma\beta}+c_{Y_{32}}a^{1+2\beta-\gamma\beta}}\\[5mm]
 %
&\iint_{\mathcal{D}_Z}\mathrm{d}\theta'\mathrm{d}\theta''\left(\frac{1}{\theta'(\theta'-\theta'')}\right)^{\beta}&&=\frac{a^{2-2 \beta }-b^{2-2 \beta }}{2 (\beta -1)^2}=\boxed{c_{Z_{11}}a^{2-2\beta}+c_{Z_{12}}b^{2-2\beta}}\\[5mm]
%
&\iint_{\mathcal{D}_Z}\mathrm{d}\theta'\mathrm{d}\theta''\left(\frac{\theta''^{2-\gamma}}{\theta'(\theta'-\theta'')}\right)^{\beta}&&=\frac{\Gamma [1-\beta ] \Gamma [-\gamma  \beta +2 \beta +1] \left(b^{2-\beta  \gamma }-a^{2-\beta  \gamma }\right)}{(\beta  \gamma -2) \Gamma [-\gamma  \beta +\beta +2]}\\[2mm]
&&&=\boxed{c_{Z_{21}} a^{2-\gamma\beta}+c_{Z_{22}}b^{2-\gamma\beta}}\\[5mm]
&\iint_{\mathcal{D}_Z}\mathrm{d}\theta'\mathrm{d}\theta''\left(\frac{\theta''^{2-\gamma}}{\theta'(\theta'-\theta'')}\right)^{\beta}\ln\left(\frac{\zeta}{\theta''^{\beta}}\right)&&=\frac{\Gamma[1+2\beta-\gamma\beta]\Gamma[1-\beta]\beta}{(\beta\gamma-2)\Gamma[2+\beta-\gamma\beta]}\left(a^{2-\gamma\beta}-b^{2-\gamma\beta}\right)\nonumber\\[2mm]
&&&\times\bigg\{H_{\beta(2-\gamma)}-H_{1+\beta-\gamma\beta}+\frac{1}{\gamma\beta-2}-\frac{1}{\beta}\log(\zeta)\nonumber\\[2mm]
&&&+\frac{a^{2-\gamma\beta}\log a-b^{2-\gamma\beta}\log b}{a^{2-\gamma\beta}-b^{2-\gamma\beta}}\bigg\}
\\[2mm]
&&&=\boxed{\begin{array}{l} a^{2-\gamma\beta}\Big(c_{Z_{31}}+c_{Z_{32}}\left(\log(a)-\frac{1}{\beta}\log(\zeta)\right)\Big)\\[2mm]
-b^{2-\gamma\beta}\Big(c_{Z_{31}}+c_{Z_{32}}\left(\log(b)-\frac{1}{\beta}\log(\zeta)\right)\Big)\end{array}}
\end{alignat}
The next step is to organise the different scalings (see Tab. \eqref{tab:terms}, where we have defined $c_{Y_i}=c_{Y_{1i}}+c_{Y_{2i}}+c_{Y_{3i}}$ and similarly for $Z$) that were found and find which is dominant. 

\begin{table}[h]
\centering
{\renewcommand{\arraystretch}{1.5}%
\setlength\tabcolsep{2pt}

\begin{tabular}{|
>{\columncolor[HTML]{C0C0C0}}l |c|c|c|c|}

\hline
 \cellcolor[HTML]{C0C0C0}&  \cellcolor[HTML]{C0C0C0}$\zeta=\left(\frac{\kappa_0}{\kappa_s}\right)^2$  &\cellcolor[HTML]{C0C0C0} $\zeta=\left(\frac{\kappa_0\kappa_c}{\kappa_s^2}\right)$ &\cellcolor[HTML]{C0C0C0} $\zeta=\left(\frac{\kappa_c}{\kappa_s}\right)^2\quad(\kappa_c\gg\kappa_s)$ &\cellcolor[HTML]{C0C0C0} $\zeta=\left(\frac{\kappa_c}{\kappa_s}\right)^2\quad(\kappa_c\leq\kappa_s)$\\ \hline
$X$ &     $\begin{array}{l}
c_{X_{11}} \left(\frac{\kappa_0}{\kappa_s}\right)^{\frac{4}{\beta}-2\gamma}+\\
c_{X_{12}}\left(\frac{\kappa_0}{\kappa_s}\right)^{2(3-\gamma)}
\end{array}$                   &    
$\begin{array}{l}
c_{X_{11}} \left(\frac{\kappa_0\kappa_c}{\kappa_s^2}\right)^{\frac{2}{\beta}-\gamma}+\\
c_{X_{12}}\left(\frac{\kappa_0\kappa_c}{\kappa_s^2}\right)^{3-\gamma}
\end{array}$                      &      \cellcolor[HTML]{000000}        & $\begin{array}{l}
c_{X_{11}} \left(\frac{\kappa_c}{\kappa_s}\right)^{\frac{4}{\beta}-2\gamma}+\\
c_{X_{12}}\left(\frac{\kappa_c}{\kappa_s}\right)^{2(3-\gamma)}
\end{array}$            \\ \hline
$Y$ &     
$\begin{array}{l}
c_{Y1} \left(\frac{\kappa_0}{\kappa_s}\right)^{\frac{4}{\beta}-2\gamma}+\\
c_{Y2}\left(\frac{\kappa_0}{\kappa_s}\right)^{4+\frac{2}{\beta}-2\gamma}
\end{array}$                      &   
$\begin{array}{l}
c_{Y1} \left(\frac{\kappa_0\kappa_c}{\kappa_s^2}\right)^{\frac{2}{\beta}-\gamma}+\\
c_{Y2}\left(\frac{\kappa_0\kappa_c}{\kappa_s^2}\right)^{2+1/\beta-\gamma}
\end{array}$                                &         \cellcolor[HTML]{000000}      &
$\begin{array}{l}
c_{Y1} \left(\frac{\kappa_c}{\kappa_s}\right)^{\frac{4}{\beta}-2\gamma}+\\
c_{Y2}\left(\frac{\kappa_c}{\kappa_s}\right)^{4+2/\beta-2\gamma}
\end{array}$\\ \hline
$Z$ &       \cellcolor[HTML]{000000}                   &    
$\begin{array}{l}
c_{Z1} \left(\frac{\kappa_0\kappa_c}{\kappa_s^2}\right)^{\frac{2}{\beta}-\gamma}+\\
\left(c_{Z_{22}}-c_{Z_{31}}\right) \left(\frac{\kappa_0}{\kappa_s}\right)^{\frac{4}{\beta}-2\gamma}+\\
\frac{c_{Z_{32}}}{\beta}\left(\frac{\kappa_0}{\kappa_s}\right)^{\frac{4}{\beta}-2\gamma}\ln\left(\frac{\kappa_c}{\kappa_0}\right)+\\
c_{Z_{12}}\left(\frac{\kappa_0\kappa_c}{\kappa_s^2}\right)^{2-\gamma}\left(\frac{\kappa_0}{\kappa_s}\right)^{\frac{4}{\beta}-4}
\end{array}$                        &      
$\begin{array}{l}
c_{Z_{11}}\pi^{2-2\beta}\left(\frac{\kappa_c}{\kappa_s}\right)^{2(2-\gamma)}+\\
\left(c_{Z_{21}}+c_{Z_{31}}\right) \pi^{2-\gamma\beta}+\\
\left(c_{Z_{22}}-c_{Z_{32}}\right) \left(\frac{\kappa_0}{\kappa_s}\right)^{\frac{4}{\beta}-2\gamma}-\\
\frac{2}{\beta}c_{Z_{32}} \pi^{2-\gamma\beta}\ln\left(\frac{\kappa_c}{\kappa_s}\right)+\\
\frac{2}{\beta}c_{Z_{32}}\left(\frac{\kappa_0}{\kappa_s}\right)^{\frac{4}{\beta}-2\gamma}\ln\left(\frac{\kappa_c}{\kappa_0}\right)+\\
c_{Z_{23}}\pi^{2-\gamma\beta}\ln(\pi)+\\
c_{Z_{12}}\left(\frac{\kappa_c}{\kappa_s}\right)^{2(2-\gamma)}\left(\frac{\kappa_0}{\kappa_s}\right)^{\frac{4}{\beta}-4}
\end{array}$ &    
$\begin{array}{l}
c_{Z1} \left(\frac{\kappa_c}{\kappa_s}\right)^{\frac{4}{\beta}-2\gamma}+\\
\left(c_{Z_{22}}-c_{Z_{31}}\right) \left(\frac{\kappa_0}{\kappa_s}\right)^{\frac{4}{\beta}-2\gamma}+\\
\frac{2c_{Z_{32}}}{\beta}\left(\frac{\kappa_0}{\kappa_s}\right)^{\frac{4}{\beta}-2\gamma}\ln\left(\frac{\kappa_c}{\kappa_0}\right)+\\
c_{Z_{12}}\left(\frac{\kappa_c}{\kappa_s}\right)^{4-2\gamma}\left(\frac{\kappa_0}{\kappa_s}\right)^{\frac{4}{\beta}-4}
\end{array}$ 
\\ \hline
\end{tabular}}
\caption{The different terms resulting from \eqref{eq:A-l3}.}
\label{tab:terms}
\end{table}

Let us note that as, the final results (Eq. \eqref{eq:A-l3}) contains $I_{\kappa_c^2/\kappa_s^2}-2I_{\kappa_0\kappa_c/\kappa_s^2}$, the terms containing $\ln(\kappa_c/\kappa_0)$ cancel. We now have many different scaling behaviours, and the question of which one dominates again depends on the value of $\beta$ as well as $\kappa_c$. As a matter of fact, if one includes the $\kappa_s^{-2}$ pre-factor in Eq. \eqref{eq:A-l3}, one recovers the same behaviour as was found for the lower bound
\begin{equation}
    I_-\leq\begin{cases}
     C_{-,1} N^{-2/\beta+\gamma-1}+C_{-,2} N^{-1}\ln N\quad &\text{if}\quad \kappa_c\gg\kappa_s\\
     N^{-1}\bigg(C_{-,3}N^{\gamma-2/\beta}+ C_{-,4}N^{(1-\alpha)(\gamma-2/\beta)}+C_{-,5} N^{(1-\alpha)(\gamma-3)}\bigg)\quad &\text{if}\quad \kappa_c\leq\kappa_s
    \end{cases}\label{eq:A-upperbound} ,
\end{equation}
where $C_{-,i}$ are constants.

This seems to go in the right direction. However, we have not explored the full integration domain yet. It turns out though that the integration domain $\mathcal{D}_s$ does not lead to any new scaling:
\begin{alignat}{6}
    I_{-,s}&=\iint\mathrm{d} x\mathrm{d} y (xy)^{-\gamma}\iint_{\mathcal{D}_s}\frac{1}{1+\dfrac{\theta'^\beta\kappa_s}{\kappa x}}\frac{1}{1+\dfrac{\theta''^\beta\kappa_s}{\kappa y}}\frac{1}{1+\dfrac{(\theta'-\theta'')^\beta}{xy}}\nonumber\\
    &\leq \iint\mathrm{d} x\mathrm{d} y (xy)^{-\gamma}\iint_{\mathcal{D}_s}1\nonumber\\
    &=\left(\frac{\kappa_0}{\kappa_s}\right)^{4/\beta} \iint\mathrm{d} x\mathrm{d} y (xy)^{-\gamma}\nonumber\\
    &=\left(\frac{\kappa_0}{\kappa_s}\right)^{4/\beta} \frac{1}{(1-\gamma)^2}\left(\left(\frac{\kappa_c}{\kappa_s}\right)^{1-\gamma}-\left(\frac{\kappa_0}{\kappa_s}\right)^{1-\gamma}\right)^2\nonumber\\
    &\simeq\frac{1}{(1-\gamma)^2}\left(\frac{\kappa_0}{\kappa_s}\right)^{2(1-\gamma+2/\beta)}\nonumber\\
    &\sim N^{-1+\gamma-2/\beta}.\label{eq:A-s}
\end{alignat}
The contribution of $\mathcal{D}_s$ is thus subleading for small $\beta$ and equally dominant as the other contributions for large $\beta$. We have thus shown that for the the upper and lower bound the dominant scaling is the same. We now have the scaling of all distinct parts, $B,A_+,A-$, so we can now combine them all. 
\begin{alignat}{6}
   \boxed{ \overline{c}\simeq
   \begin{cases}
    C_1 N^{2-2/\beta}+C_2 N^{2-\gamma}\ln N\quad &\text{if} \quad \kappa_c\gg\kappa_s\\
    C_{3}N^{2-2/\beta}+ C_{4}N^{2-2/\beta-\alpha(\gamma-2/\beta)}+C_{5} N^{-1-\alpha(\gamma-3)}\quad &\text{if}\quad \kappa_c\leq\kappa_s
   \end{cases}}
\end{alignat}
Let us discuss the limiting cases of $\alpha$. When $\alpha=0$, $\kappa_0\sim\kappa_c$, and the network thus has a homogeneous degree distribution. Then, $C\simeq (C_3+C_4)N^{2-2/\beta}+C_5N^{-1}$. If $\alpha=1$, i.e. $\kappa_c\sim\kappa_s$, the scaling becomes $C\simeq C_3N^{2-2/\beta}+(C_4+C_5)N^{2-\gamma}$. 

\subsubsection{Case $\beta=1$}

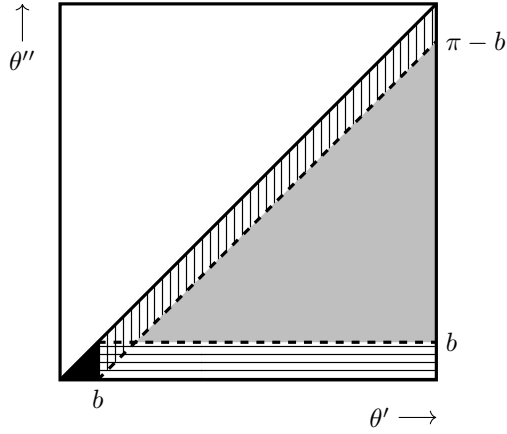
\begin{figure}[h]
	\centering
	\begin{tikzpicture}
		\def\AA{2.5}
		\fill[gray!50] (-0.6*\AA,-0.8*\AA) -- (1*\AA,-0.8*\AA) -- (1*\AA,0.8*\AA) -- cycle;
		\fill[black] (-1*\AA,-1*\AA) -- (-0.8*\AA,-1*\AA) -- (-0.8*\AA,-0.8*\AA) --cycle;
		\fill[pattern=horizontal lines] (-0.8*\AA,-1*\AA) -- (1*\AA,-1*\AA) -- (\AA,-0.8*\AA) -- (-0.8*\AA,-0.8*\AA) -- cycle;
		\fill[pattern=vertical lines] (-0.8*\AA,-1*\AA) -- (1*\AA,0.8*\AA) -- (\AA,\AA) -- (-0.8*\AA,-0.8*\AA) -- cycle;
		\draw[black, very thick] (-1*\AA,-1*\AA) rectangle (1*\AA,1*\AA);
		\draw[black,very thick] (-1*\AA,-1*\AA) -- (1*\AA,1*\AA);
		\draw[black,very thick] (-0.8*\AA,-1*\AA) -- (-0.8*\AA,-0.8*\AA);
		\draw[dashed,very thick] (-0.8*\AA,-0.8*\AA) -- (1*\AA,-0.8*\AA);
		\draw[dashed,very thick] (-0.8*\AA,-1*\AA) -- (1*\AA,0.8*\AA);
		\draw[->] (-1.2*\AA,0.8*\AA) -- (-1.2*\AA,1*\AA);
		\node at (-1.2*\AA,0.7*\AA) { $\theta''$}; 
		\draw[->] (0.8*\AA,-1.2*\AA) -- (1*\AA,-1.2*\AA);
		\node at (0.7*\AA,-1.2*\AA) { $\theta'$};
		\node[above] at (-0.8*\AA,-1.2*\AA) { $b$};
		\node[right] at (1.*\AA,0.8*\AA) { $\pi-b$};
		\node[right] at (1.*\AA,-0.8*\AA) { $b$};
	\end{tikzpicture}%
	
	\caption{Integration regions where $b=\frac{\kappa_0^2}{\kappa_s^2}$ The black region is region $\mathcal{D}_s$. The horizontally striped region is region $\mathcal{D}_2$. The vertically striped region is region $\mathcal{D}_3$. The grey region is region $\mathcal{D}_4$.}
	\label{fig:integrationregions2}
\end{figure}

We now turn to the limit $\beta=1$. The general practice of finding upper and lower bounds for the various relevant integrals will be again pursued here, and in many cases the integrals examined will be similar to the ones studied above. However, there are some important differences that force us to treat this case separately. For one, we know that in the case of $\beta=1$, $\mu$ scales as $\hat{\mu}\sim (\ln N)^{-1}$ instead of $\hat{\mu}\sim N^{1-\beta}$, and thus $\kappa_s\sim\sqrt{N\ln N}$, which of course alters scaling. We will represent all integrals evaluated at $\beta=1$ by a tilde ($\tilde{A}_-,\tilde{A}_+,\tilde{B}$). We start with $\tilde{B}$:
\begin{alignat}{6}
    \tilde{B}&=\pi\left(\frac{\kappa_c}{\kappa_s}\right)^{1-\gamma}\int\limits^1_0\mathrm{d}\theta\int\limits^1_{\kappa_0/\kappa_c}\mathrm{d}x\frac{x^{-\gamma}}{1+\dfrac{\pi\theta\kappa_s^2}{\kappa_c\kappa x}}\nonumber\\
    &=\pi\left(\frac{\kappa_c}{\kappa_s}\right)^{1-\gamma}\bigg\{\frac{\kappa\kappa_c}{\pi\kappa_s^2}\frac{\log\left(1+\frac{\pi\kappa_s^2}{\kappa_c\kappa}\right)}{2-\gamma}+\frac{1}{\gamma-2}\left(\frac{\kappa_0}{\kappa_c}\right)^{2-\gamma}\frac{\kappa\kappa_c}{\pi\kappa_s^2}\log\left(1+\frac{\pi\kappa_s^2}{\kappa\kappa_0}\right)\nonumber\\[2mm]
    &+\frac{1}{(\gamma-2)(\gamma-1)}\left({}_2F_1\left[\begin{array}{c}1,\gamma-1\\\gamma\end{array};-\frac{\pi\kappa_s^2}{\kappa\kappa_c}\right]-\left(\frac{\kappa_0}{\kappa_c}\right)^{1-\gamma}{}_2F_1\left[\begin{array}{c}1,\gamma-1\\\gamma\end{array};-\frac{\pi\kappa_s^2}{\kappa\kappa_0}\right]\right)\bigg\}.
\end{alignat}
The second term is dominant and thus $\tilde{B}$ scales as
\begin{alignat}{6}
    \tilde{B}\sim\kappa_s^{\gamma-3}\log(\kappa_s)\sim N^{\frac{\gamma-3}{2}}(\log N)^{\frac{\gamma-1}{2}}.
\end{alignat}
For the lower bound of the numerator of the clustering coefficient we can use the result found in Eq. \eqref{eq:A+lowerbound} as nowhere was it assumed that $\beta<1$. Irrespective of $\kappa_c$ this gives us
\begin{equation}
    \tilde{A}_{+}\leq \tilde{c}_{+}N^{\gamma-3}\left(\ln N\right)^{\gamma-3}.
\end{equation}
For the upper bound of $\tilde{A}_{-}$ we cannot follow the same path as was done in the case of general $\beta$. This is because the upper bound employed, given by Eq. \eqref{eq:A-l3}, diverges in the $\beta=1$ limit. Thus, we must find a stricter bound. This is done by once again dividing the angular integration domain, this time in four pieces: $\mathcal{D}_{s}=[0,(\kappa_0/\kappa_s)^2]\times[0,\theta']$, $\mathcal{D}_{2}=[(\kappa_0/\kappa_s)^2,\pi]\times[0,(\kappa_0/\kappa_s)^2]$, $\mathcal{D}_{3}=[(\kappa_0/\kappa_s)^2,\pi]\times[\theta'-(\kappa_0/\kappa_s)^2,\theta']$ and $\mathcal{D}_{3}=[2(\kappa_0/\kappa_s)^2,\pi]\times[(\kappa_0/\kappa_s)^2,\theta'-(\kappa_0/\kappa_s)^2]$, as represented in Fig. \ref{fig:integrationregions2}.
Note that regions $\mathcal{D}_2$ and $\mathcal{D}_3$ overlap, but that is not a problem as our integrand is positive and counting a region double just increases the value of the integral, which in turn work for our purposes as we are only looking for an upper bound. For the region $\mathcal{D}_s$ we can use the result \eqref{eq:A-s}:
\begin{equation}
    \tilde{A}_{-,s}\leq \tilde{c}_{-,s}N^{\gamma-3}\left(\ln N\right)^{\gamma-3}.
\end{equation}
Turning to $\mathcal{D}_2$ we obtain
\begin{alignat}{6}
    \tilde{A}_{-,2}&=\iint\mathrm{d} x\mathrm{d} y (xy)^{-\gamma}\iint_{\mathcal{D}_2}\frac{\mathrm{d}\theta'\mathrm{d}\theta''}{1+\dfrac{\theta'\kappa_s}{\kappa x}}\frac{1}{1+\dfrac{\theta''\kappa_s}{\kappa y}}\frac{1}{1+\dfrac{\theta'-\theta''}{xy}}\nonumber\\
    &\leq\frac{\kappa}{\kappa_s}\iint_{\mathcal{D}_2}\frac{\mathrm{d}\theta'\mathrm{d}\theta''}{\theta'}\int\limits_{\kappa_s/\kappa_c}^{\kappa_s/\kappa_0}\mathrm{d}x\int\limits_{\kappa_s/\kappa_c}^{\kappa_s/\kappa_0}\mathrm{d}y\frac{x^{\gamma-3}y^{\gamma-2}}{1+xy(\theta'-\theta'')}
\end{alignat}
where we have bounded the integral by decreasing the size of the denominators of the first and second terms. We also performed a change of variables of $x$ and $y$. We now extend the lower bounds of the $x$ and $y$ integrals to zero, which can be done as our integral is positive, and so the resulting integral will be larger or equal to the original one. 
\begin{alignat}{6}
    \tilde{A}_{-,2}&\leq\frac{\kappa}{\kappa_s}\iint_{\mathcal{D}_2}\frac{\mathrm{d}\theta'\mathrm{d}\theta''}{\theta'}\int\limits_{0}^{\kappa_s/\kappa_0}\mathrm{d}x\int\limits_{0}^{\kappa_s/\kappa_0}\mathrm{d}y\frac{x^{\gamma-3}y^{\gamma-2}}{1+xy(\theta'-\theta'')}\nonumber\\
    &=\frac{\kappa}{\kappa_s}(\kappa_s/\kappa_0)^{2\gamma-3}\iint_{\mathcal{D}_2}\frac{\mathrm{d}\theta'\mathrm{d}\theta''}{\theta'}\bigg(\Phi\left[-\frac{\kappa_s^2}{\kappa_0^2}(\theta'-\theta''),1,\gamma-2\right]-\Phi\left[-\frac{\kappa_s^2}{\kappa_0^2}(\theta'-\theta''),1,\gamma-1\right]\bigg)
\end{alignat}
We know again have the situation that depending on the values of the angular coordinates, the arguments of the $\Phi$'s diverge or go to zero. For the region $\mathcal{D}_{2s}=[b,2b]\times[0,b]$, $\theta'-\theta''\in[0,b]$, so the argument lies between zero and one. For the region $\mathcal{D}_{2l}=[2b,\pi]\times[0,b]$, $\theta'-\theta''\in[b,\pi]$, so the argument is larger than one. We first turn to the second region. Here the argument can diverge and we should thus perform a similar transformation as Eq. \eqref{eq:LerchPhiTransform}. It is not exactly the same as the second argument of the $\Phi$'s is now $1$ and not two $2$, but the derivation is equivalent. This leads us to
\begin{alignat}{6}
&\frac{\kappa}{\kappa_s}(\kappa_s/\kappa_0)^{2\gamma-3}\iint_{\mathcal{D}_{2l}}\frac{\mathrm{d}\theta'\mathrm{d}\theta''}{\theta'}&&\bigg(&&\Phi\left[-\frac{\kappa_s^2}{\kappa_0^2}(\theta'-\theta''),1,\gamma-2\right]-\Phi\left[-\frac{\kappa_s^2}{\kappa_0^2}(\theta'-\theta''),1,\gamma-1\right]\bigg)\nonumber\\
&=\frac{\kappa}{\kappa_s}(\kappa_s/\kappa_0)^{2\gamma-3}\iint_{\mathcal{D}_{2l}}\frac{\mathrm{d}\theta'\mathrm{d}\theta''}{\theta'}&&\bigg(&&\left(\frac{\kappa_s^2}{\kappa_0^2}(\theta'-\theta'')\right)^{2-\gamma}\big(\Phi[-1,1,3-\gamma]+\Phi[-1,1,2-\gamma]\big)\nonumber\\
&&&-&&\left(\frac{\kappa_s^2}{\kappa_0^2}(\theta'-\theta'')\right)^{-1}\Phi\left[-\left(\frac{\kappa_s^2}{\kappa_0^2}(\theta'-\theta'')\right)^{-1},1,3-\gamma\right]\nonumber\\
&&&+&&\left(\frac{\kappa_s^2}{\kappa_0^2}(\theta'-\theta'')\right)^{1-\gamma}\big(\Phi[-1,1,2-\gamma]+\Phi[-1,1,1-\gamma]\big)\nonumber\\
&&&-&&\left(\frac{\kappa_s^2}{\kappa_0^2}(\theta'-\theta'')\right)^{-1}\Phi\left[-\left(\frac{\kappa_s^2}{\kappa_0^2}(\theta'-\theta'')\right)^{-1},1,2-\gamma\right]\bigg)\nonumber\\
&\leq\frac{\kappa}{\kappa_s}(\kappa_s/\kappa_0)^{2\gamma-3}\iint_{\mathcal{D}_{2l}}\frac{\mathrm{d}\theta'\mathrm{d}\theta''}{\theta'}&&\bigg(&&\left(\frac{\kappa_s^2}{\kappa_0^2}(\theta'-\theta'')\right)^{2-\gamma}\big(\Phi[-1,1,3-\gamma]+\Phi[-1,1,2-\gamma]\big)\nonumber\\
&&&+&&\left(\frac{\kappa_s^2}{\kappa_0^2}(\theta'-\theta'')\right)^{1-\gamma}\big(\Phi[-1,1,2-\gamma]+\Phi[-1,1,1-\gamma]\big)\nonumber\\
&&&-2&&\left(\frac{\kappa_s^2}{\kappa_0^2}(\theta'-\theta'')\right)^{-1}\bigg)\sim\kappa_s^{2(\gamma-3)}\sim N^{\gamma-3}\left(\ln N\right)^{\gamma-3}.
\end{alignat}
For $\mathcal{D}_{2s}$ we can immediately bound away the $\Phi$ to find
\begin{alignat}{6}
&\frac{\kappa}{\kappa_s}(\kappa_s/\kappa_0)^{2\gamma-3}\iint_{\mathcal{D}_{2s}}\frac{\mathrm{d}\theta'\mathrm{d}\theta''}{\theta'}\bigg(\Phi\left[-\frac{\kappa_s^2}{\kappa_0^2}(\theta'-\theta''),1,\gamma-2\right]-\Phi\left[-\frac{\kappa_s^2}{\kappa_0^2}(\theta'-\theta''),1,\gamma-1\right]\bigg)\nonumber\\
	&\leq\frac{\kappa}{\kappa_s}(\kappa_s/\kappa_0)^{2\gamma-3}\iint_{\mathcal{D}_{2s}}\frac{\mathrm{d}\theta'\mathrm{d}\theta''}{\theta'}=\frac{\kappa}{\kappa_s}(\kappa_s/\kappa_0)^{2\gamma-5}\ln{2}\sim N^{\gamma-3}\left(\ln N\right)^{\gamma-3}.
\end{alignat}
Combining the two results we find that $\tilde{A}_{-,2}\leq \tilde{c}_{-,2} N^{\gamma-3}\left(\ln N\right)^{\gamma-3}$ as expected. 
Then we investigate to $\mathcal{D}_3$:
\begin{alignat}{6}
    \tilde{A}_{-,3}&=\iint\mathrm{d} x\mathrm{d} y (xy)^{-\gamma}\iint_{\mathcal{D}_3}\mathrm{d}\theta'\mathrm{d}\theta''\frac{1}{1+\dfrac{\theta'\kappa_s}{\kappa x}}\frac{1}{1+\dfrac{\theta''\kappa_s}{\kappa y}}\frac{1}{1+\dfrac{\theta'-\theta''}{xy}}\nonumber\\
    &=\iint\mathrm{d} x\mathrm{d} y (xy)^{-\gamma}\iint_{\mathcal{D}_2}\mathrm{d}\theta'\mathrm{d}\theta''\frac{1}{1+\dfrac{\theta'\kappa_s}{\kappa x}}\frac{1}{1+\dfrac{(\theta'-\theta'')\kappa_s}{\kappa y}}\frac{1}{1+\dfrac{\theta''}{xy}}\nonumber\\
    &\leq\left(\frac{\kappa}{\kappa_s}\right)^2\iint\mathrm{d} x\mathrm{d} y x^{1-\gamma}y^{1-\gamma}\iint_{\mathcal{D}_2}\mathrm{d}\theta'\mathrm{d}\theta''\frac{1}{\theta'}\frac{1}{\theta'-\theta''}\nonumber\\
    &=\frac{1}{(2-\gamma)^2}\left(\frac{\kappa}{\kappa_s}\right)^2\left(\frac{\kappa_0}{\kappa_s}\right)^{2(2-\gamma)}\bigg(\frac{\pi^2}{6}-\text{Li}_2\left[\frac{\kappa_0^2}{\kappa_s^2\pi}\right]\bigg)\nonumber\\[2mm]
    &\sim\kappa_s^{2(\gamma-3)}\sim N^{\gamma-3}\left(\ln N\right)^{\gamma-3}.
\end{alignat}
Here $\text{Li}_2(z)$ is the dilogarithm. The final region to be studied is $\mathcal{D}_4$:
\begin{alignat}{6}
    \tilde{A}_{-,4}&=\iint\mathrm{d} x\mathrm{d} y (xy)^{-\gamma}\iint_{\mathcal{D}_4}\mathrm{d}\theta'\mathrm{d}\theta''\frac{1}{1+\dfrac{\theta'\kappa_s}{\kappa x}}\frac{1}{1+\dfrac{\theta''\kappa_s}{\kappa y}}\frac{1}{1+\dfrac{\theta'-\theta''}{xy}}\nonumber\\
    &\leq\left(\frac{\kappa}{\kappa_s}\right)^2\iint\mathrm{d} x\mathrm{d} y (xy)^{1-\gamma}\iint_{\mathcal{D}_4}\mathrm{d}\theta'\mathrm{d}\theta''\frac{1}{\theta'\theta''}\frac{1}{1+\frac{\theta'-\theta''}{xy}}\nonumber\\
    &=\left(\frac{\kappa}{\kappa_s}\right)^2\int\limits^{\kappa_s/\kappa_0}_{\kappa_s/\kappa_c}\mathrm{d} x\int\limits^{\kappa_s/\kappa_0}_{\kappa_s/\kappa_c}\mathrm{d} y (xy)^{\gamma-3}\iint_{\mathcal{D}_4}\mathrm{d}\theta'\mathrm{d}\theta''\frac{1}{\theta'\theta''}\frac{1}{1+xy(\theta'-\theta'')}\nonumber\\
    &\leq\left(\frac{\kappa}{\kappa_s}\right)^2\int\limits^{\kappa_s/\kappa_0}_0\mathrm{d} x\int\limits^{\kappa_s/\kappa_0}_{0}\mathrm{d} y (xy)^{\gamma-3}\iint_{\mathcal{D}_4}\mathrm{d}\theta'\mathrm{d}\theta''\frac{1}{\theta'\theta''}\frac{1}{1+xy(\theta'-\theta'')}\nonumber\\
    &=\left(\frac{\kappa}{\kappa_s}\right)^2\left(\frac{\kappa_s}{\kappa_0}\right)^{2(\gamma-2)}\iint_{\mathcal{D}_4}\mathrm{d}\theta'\mathrm{d}\theta''\frac{1}{\theta'\theta''}\Phi\left[-\frac{\kappa_s^2}{\kappa_0^2}(\theta'-\theta''),2,\gamma-2\right]\nonumber\\
    &\leq\left(\frac{\kappa}{\kappa_s}\right)^2\iint_{\mathcal{D}_4}\mathrm{d}\theta'\mathrm{d}\theta''\frac{1}{\theta'\theta''}\bigg\{(\theta'-\theta'')^{2-\gamma}\left[\Psi(.)\log\left(\frac{\kappa_s^2}{\kappa_0^2}(\theta'-\theta'')\right)+\vartheta(.)\right]\nonumber\\
    &+\left(\frac{\kappa_s}{\kappa_0}\right)^{2(\gamma-3)}(\theta'-\theta'')^{-1}(3-\gamma)^{-2}\bigg\}.\label{eq:A-b1S4}
\end{alignat}
Let us investigate the term with the logarithm first. 
\begin{alignat}{6}
    &\left(\frac{\kappa}{\kappa_s}\right)^2\int\limits^{\pi}_{2\kappa_0^2/\kappa_s^2}\mathrm{d}\theta'\int\limits^{\theta'-\kappa_0^2/\kappa_s^2}_{\kappa_0^2/\kappa_s^2}\mathrm{d}\theta''\frac{(\theta'-\theta'')^{2-\gamma}}{\theta'\theta''}\log\left(\frac{\kappa_s^2}{\kappa_0^2}(\theta'-\theta'')\right)\nonumber\\
    =&\left(\frac{\kappa}{\kappa_s}\right)^2\left(\frac{\kappa_0}{\kappa_s}\right)^{2(2-\gamma)}\int\limits^{\pi\kappa_s^2/\kappa_0^2}_{2}\mathrm{d}\theta'\int\limits^{\theta'-1}_{1}\mathrm{d}\theta''\frac{(\theta'-\theta'')^{2-\gamma}}{\theta'\theta''}\log\left(\theta'-\theta''\right)\nonumber\\
    =&\left(\frac{\kappa}{\kappa_s}\right)^2\left(\frac{\kappa_0}{\kappa_s}\right)^{2(2-\gamma)}\int\limits^{\pi\kappa_s^2/\kappa_0^2}_{2}\mathrm{d}\theta'\int\limits^{\theta'-1}_{1}\mathrm{d}\theta''\frac{(\theta'')^{2-\gamma}}{\theta'(\theta'-\theta'')}\log\left(\theta''\right).
\end{alignat}
This can then be evaluated. The $\theta''$ integral leads to a variety of different terms, which need to be treated separately. Some variable transformations need to be performed, and some special functions need to be expanded to their series representation. It can be shown that the integral to leading order is constant in $N$, implying that the logarithm term of $\tilde{A}_{-,4}$ scales as $\kappa_s^{2(\gamma-3)}$. The other two terms in expression \eqref{eq:A-b1S4} are easier to evaluate:
\begin{alignat}{6}
    &\iint_{\mathcal{D}_4}\mathrm{d}\theta'\mathrm{d}\theta''\frac{1}{\theta'\theta''}(\theta'-\theta'')^{2-\gamma}&&=\frac{\left(b^{2-\gamma}+\pi^{2-\gamma}\right)}{\gamma-2}\bigg\{B_{1-\frac{b}{\pi}}[3-\gamma,\gamma-2]-B_{\frac{1}{2}}[3-\gamma,\gamma-2]\bigg\} \nonumber\\[2mm]
    &&&+\frac{b^{2-\gamma}\ln\left(2-\frac{2b}{\pi}\right)}{\gamma-2}\sim \left(\frac{\kappa_0}{\kappa_s}\right)^{2(2-\gamma)}\\
    &\iint_{\mathcal{D}_4}\mathrm{d}\theta'\mathrm{d}\theta''\frac{1}{\theta'\theta''}(\theta'-\theta'')^{-1}&&=\frac{2 \log \left(2-\frac{2 b}{\pi }\right)}{b}-\frac{2 \log \left(\frac{\pi }{b}-1\right)}{\pi }\sim \frac{\kappa_s^2}{\kappa_0^2}.
\end{alignat}
Plugging this back in we find that also the integral over the region $\mathcal{D}_4$ scales as $N^{\gamma-3}(\ln N)^{\gamma-3}$.\\

Thus, we can finally conclude that for $\beta=1$, the clustering coefficient must scale as
\begin{equation}
    \boxed{\overline{c}\sim\frac{N^{\gamma-3}(\log N)^{\gamma-3}}{N^{\gamma-3}(\log N)^{\gamma-1}}=(\log N)^{-2}}.
\end{equation}
With this we have found the critical exponent $\eta/\nu=2$.
\newpage

\subsection{Exponent $\eta$}\label{sec:exponenteta}
In this section we show that the scaling exponent $\eta$ that encodes how the clustering approaches zero when $\beta\rightarrow \beta_c^+=1$. As this only requires working on the low temperature side of the transition, we can directly work in the thermodynamic limit (we thus take first the limit $N\rightarrow\infty$ and then $\beta\rightarrow1$). To this end, we denote the general definition of the clustering coefficient with hidden degree $\kappa$ and (without loss of generality) spacial coordinate $r=0$
\begin{alignat}{6}
    \overline{c}&(\kappa)=
    &\frac{\int\limits_{\kappa_0}^\infty\mathrm{d}\kappa'\int\limits_{\kappa_0}^\infty\mathrm{d}\kappa''\int\limits_{-\infty}^{\infty}\mathrm{d}r'\int\limits_{-\infty}^{\infty}\mathrm{d}r''\rho(\kappa')\rho(\kappa'')p(\kappa,\kappa',|r'|)p(\kappa,\kappa'',|r''|)p(\kappa',\kappa'',|r'-r''|)}{\left(\int\limits_{\kappa_0}^\infty\mathrm{d}\kappa'\int\limits_{-\infty}^{\infty}\mathrm{d}r'\rho(\kappa')p(\kappa,\kappa',|r'|)\right)^2}.
\end{alignat}
where we can use connection probability \eqref{eq:connectionprobabilityThermolimit} and $\hat{\mu}$ \eqref{eq:mulowtemp}.

Let us first turn to the denominator:
\begin{alignat}{6}
    \int\mathrm{d}\kappa'\rho(\kappa')\int\limits_{-\infty}^\infty\frac{\mathrm{d}r'}{1+\left(\frac{r'}{\kappa\kappa' \hat{\mu}}\right)^{\beta}}=\kappa,
\end{alignat}
where we have plugged in the definition of $\hat{\mu}$ and used that $\langle k\rangle=\frac{\gamma-1}{\gamma-2}\kappa_0$. \\

The next step is the numerator. We first perform the transformation $t=r'/(\kappa\kappa'\hat{\mu})$ and $\tau=r''/(\kappa\kappa''\hat{\mu})$ to obtain
\begin{alignat}{6}
    \overline{c}(\kappa)=\frac{\hat{\mu}^2}{4}(\gamma-1)^2\kappa_0^{2\gamma-2}\iiiint\mathrm{d}\kappa'\mathrm{d}\kappa''\mathrm{d}t\mathrm{d}\tau\frac{(\kappa'\kappa'')^{1-\gamma}}{1+|t|^{\beta}}\frac{1}{1+|\tau|^{\beta}}\frac{1}{1+\left|\frac{\kappa t}{\kappa''}-\frac{\kappa \tau}{\kappa'}\right|^{\beta}}.
\end{alignat}
We know that $\hat{\mu}^2\sim(\beta-1)^2$. This is exactly the scaling that we expect from numerical investigation for the clustering coefficient. Thus, all we need to prove is that at $\beta=1$, the numerator is finite. If so, its $(\beta-1)$ dependence must be order $\mathcal{O}(1)$. If the full expression contained $(\beta-1)^{-n}$ terms with $n>0$ it would diverge at the critical point and if the dominant term was $\mathcal{O}((\beta-1)^n)$ with $n>0$ the numerator would go to zero at the critical point. And indeed, numerical integration shows that at $\beta=1$ the numerator is finite, leading to the conclusion that 
\begin{equation}
    \boxed{\overline{c}(\kappa)\sim(\beta-1)^2}
\end{equation}
such that $\eta=2$, which in turn implies that $\nu=1$.

\newpage
\section{Real Networks}
As was stated in the main text, the DPG algorithm can be used to find the temperature of its embedding in the $\mathbb{S}_1$ model. We list here a collection of real networks and their corresponding inverse temperatures. We choose to restrict ourselves to models where the inverse temperature lies below or close to the transition point $\beta_c$. 

\begin{table}[H]
\centering
\begin{tabular}{|l|r|c|c|c|c|c|c|}
\hline
 Network Names   & Type   & $|V|$ & $|E|$ & $\langle k\rangle$ & Target $\overline{c}$& $\beta$ \\ \hline
 CElegans-C~\cite{Ahn2006}& Biological - Brain   & 279 & 2287 & 16& 0.34& 1.5\\ 
 Drosophila1-C~\cite{Takemura2013} \ &Biological - Brain   & 350 &2887 & 16& 0.25& 1.1\\
 Drosophila2-C~\cite{Takemura2013} &Biological - Brain   & 1770 &8905 & 10& 0.33& 1.1 \\ 
 Arabidopsis-G~\cite{AIMConsortium2011} & Biological - Cell   & 4519 & 10721 &  4.7& 0.16&1.2  \\
 CElegans-G~\cite{Ahn2006} & Biological - Cell   & 3692 & 7650 & 4.2& 0.11 & 0.77\\
 Drosophila-G~\cite{Stark2006}&Biological - Cell    & 8114 & 38909 & 9.6& 0.12& 1.1\\
 Human1-P~\cite{Chang2014} & Biological - Cell   & 913 & 7472 & 16& 0.23& 1.0\\
 Human2-P~\cite{Chang2014} & Biological - Cell   & 1090 & 9369 & 17& 0.20 & 1.0\\
 Mus-G~\cite{Stark2006} & Biological - Cell   &7402 & 16858 & 4.6& 0.13 & 1.1\\
 %
 %
 Rattus-G~\cite{Stark2006} & Biological - Cell   & 2350 & 3484 & 3.0& 0.22 & 0.74\\
 Yeast1-P~\cite{Yu2008} & Biological - Cell   & 1647 & 2518 & 3.1& 0.10 & 1.2\\
 Yeast2-P~\cite{Yeast2007} &Biological - Cell   & 1458 & 1948 & 2.7& 0.14 & 1.5\\
 Polblogs-H~\cite{Adamic2005} & Citation - Hyperlinks   & 1222 & 16714 & 27& 0.36 & 1.1\\
 Wiki-H~\cite{Heaberlin2016} & Citation - Hyperlinks  & 1872 & 15367 & 16& 0.42 & 1.3 \\
 Ecological~\cite{Dunne2014} & Ecological - Troffic   & 700 & 6495 & 18 & 0.10& 0.15\\
 Commodities~\cite{Grady2012}  & Economic - Commodities  & 374 & 1090 & 5.8 & 0.22& 1.2\\ 
 Friends-OFF~\cite{Moody2001} & Social Offline - Friends   & 2539 & 10455 & 8.2& 0.15 & 1.4\\ 
 Airports1~\cite{Opsahl2011} & Transport - Flights   & 1572 & 17214 & 22 & 0.64& 1.4\\\hline
\end{tabular}
\caption{Properties of a selection of networks with the inverse temperature $\beta$ obtained with the DPG algorithm. Only networks with $\beta<1.5$ are shown. }
\label{tabRN}
\end{table}
\newpage
\section{Supplementary Figures}

\begin{figure}[t]
	\vspace{-8mm}
	\centering
	\includegraphics[width=0.75\linewidth]{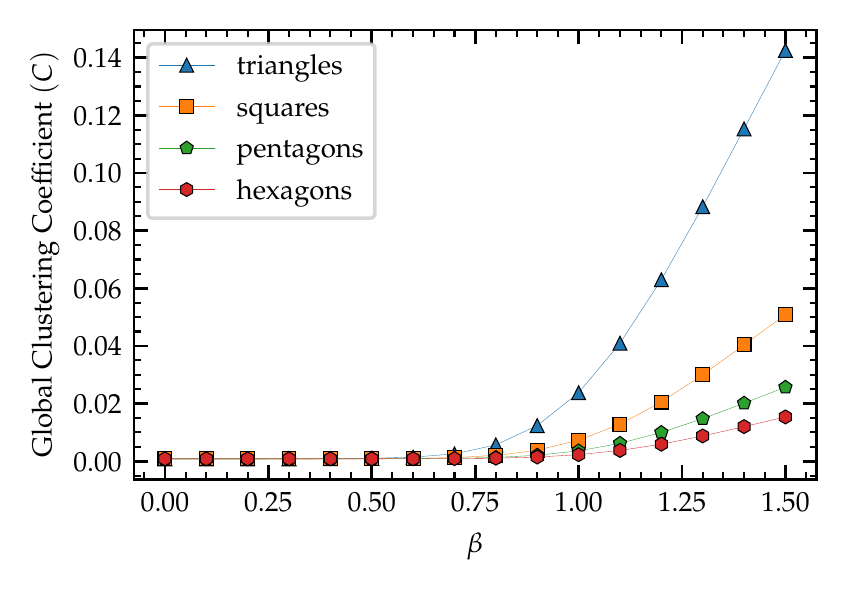}
	\vspace{-0.2cm}\caption{The global clustering coefficient for different sized chordless cycles as a function of the inverse temperature $\beta$. The results shown are for networks of size $N=5000$ and $\langle k\rangle=6$}
	\label{fig:HOcycles}
\end{figure}

\begin{figure}[t]
	\vspace{-1mm}
	\centering
	\begin{subfigure}[b]{0.325\textwidth}
		\includegraphics[width=1\textwidth]{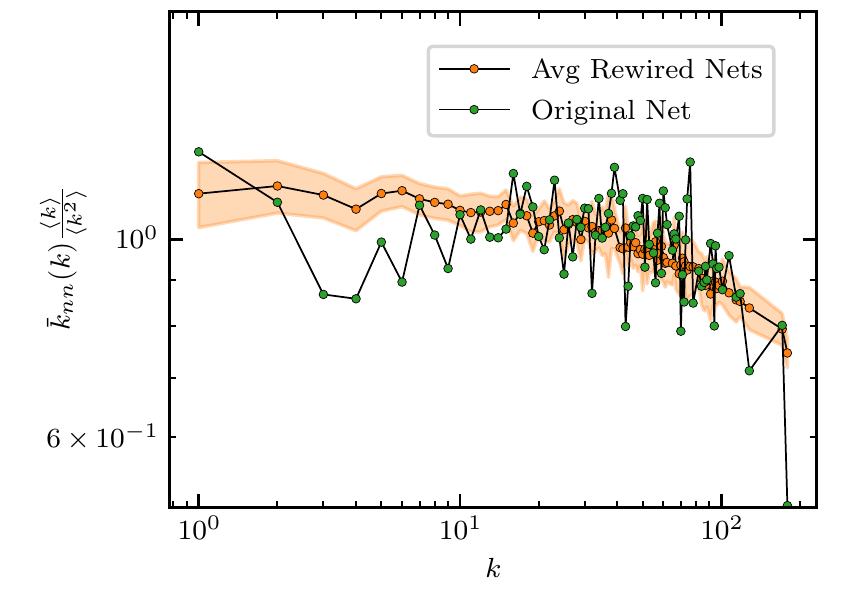}
		\caption{}
		\label{fig:Ileumknn} 
	\end{subfigure}
	\hfill
	\begin{subfigure}[b]{0.325\textwidth}
		\includegraphics[width=1\textwidth]{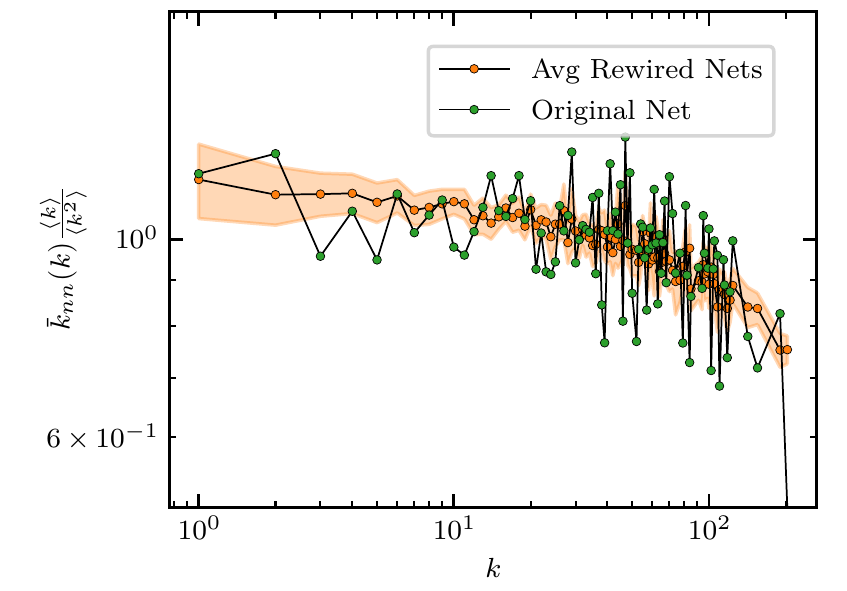}
		\caption{}
		\label{fig:Toothknn}
	\end{subfigure}
	\hfill
	\begin{subfigure}[b]{0.325\textwidth}
		\includegraphics[width=1\textwidth]{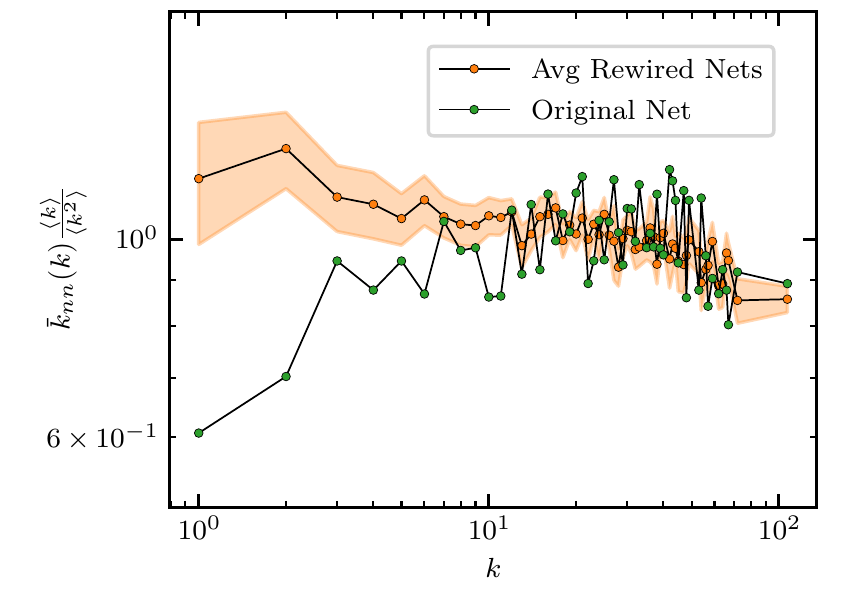}
		\caption{}
		\label{fig:Drosophilaknn}
	\end{subfigure}
	
	\begin{subfigure}[b]{0.325\textwidth}
	\includegraphics[width=1\textwidth]{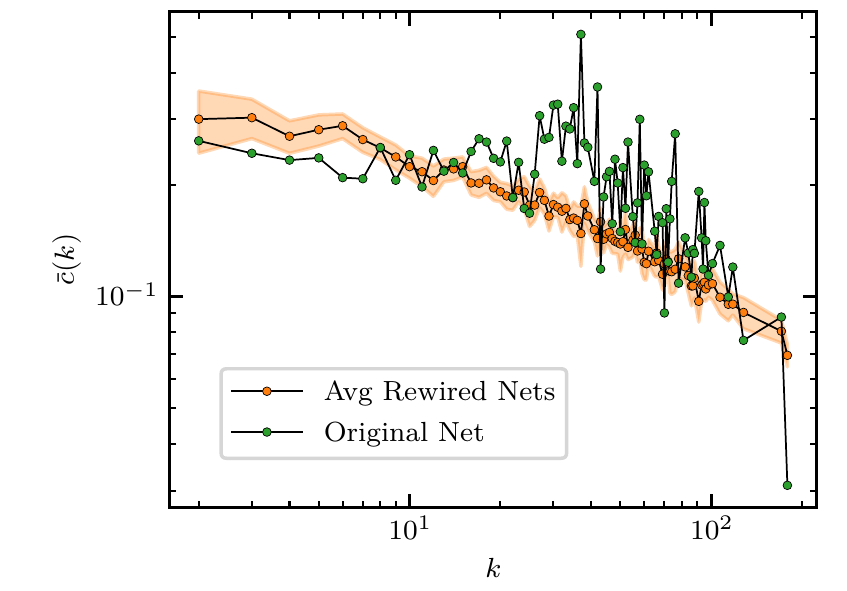}
	\caption{}
	\label{fig:Ileumck} 
	\end{subfigure}
	\hfill
	\begin{subfigure}[b]{0.325\textwidth}
		\includegraphics[width=1\textwidth]{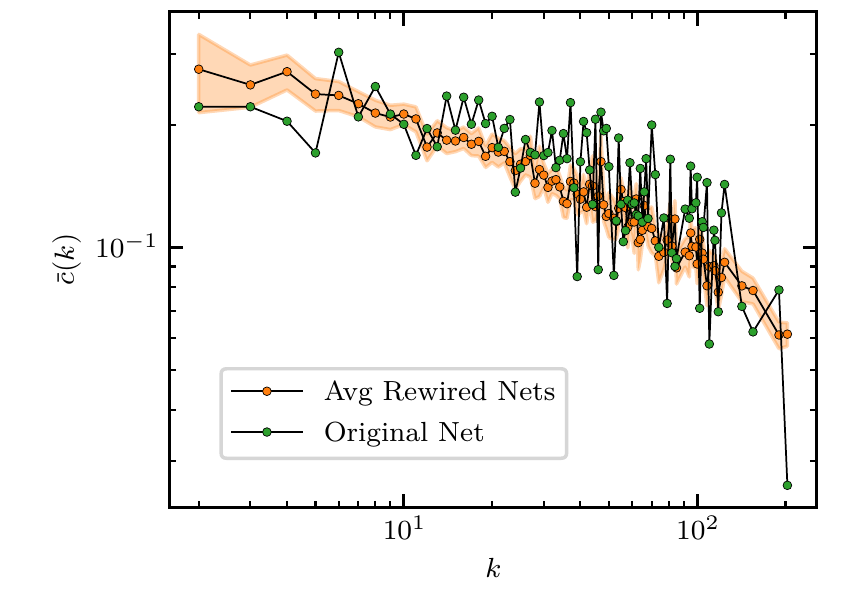}
		\caption{}
		\label{fig:Toothck}
	\end{subfigure}
	\hfill
	\begin{subfigure}[b]{0.325\textwidth}
		\includegraphics[width=1\textwidth]{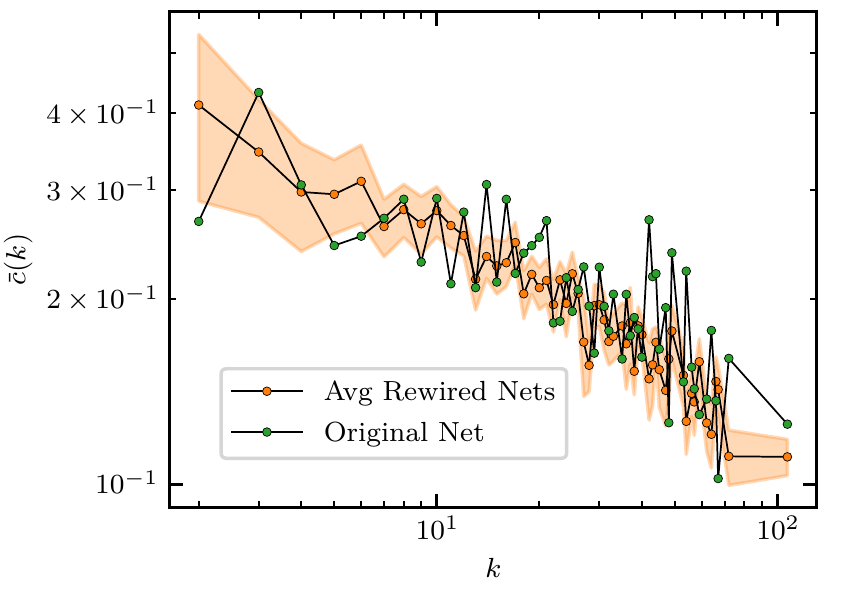}
		\caption{}
		\label{fig:Drosophilack}
	\end{subfigure}
	
	\caption{Degree-degree correlations (upper row) and average clustering coefficient per degree (lower row) for three of the real networks in Tab. \ref{tabRN}. The first column corresponds to the Human1-P network, the second to Human2-P and the third to Drosophila-G. The green points represent the network measures corresponding to the original network. The orange points represent the the average of 100 randomized networks at the $\beta$ that reproduces the correct global clustering coefficient (see Tab. \ref{tabRN}). }
	\label{fig:DynamicsDR}
\end{figure}

\clearpage

\bibliographystyle{nature}
\bibliography{geometry3}